\documentclass[10pt]{article}
\usepackage{latexsym}
\usepackage{graphicx}
\usepackage{epsfig}

\title{Set-Theoretic Preliminaries for Computer Scientists
}
\author{M.H. van Emden}
\date{\small{
        Research Report DCS-304-IR \\
        Department of Computer Science, University of Victoria
      }
}

\begin{document}
\maketitle

\begin{abstract}
The basics of set theory are usually copied,
directly or indirectly,
by computer scientists from introductions to mathematical texts.
Often mathematicians are content with special cases
when the general case is of no mathematical interest.
But sometimes what is of no mathematical interest
is of great practical interest in computer science.
For example, non-binary relations in mathematics
tend to have numerical indexes
and tend to be unsorted.
In the theory and practice of relational databases
both these simplifications are unwarranted.
In response to this situation
we present here an alternative to the ``set-theoretic preliminaries''
usually found in computer science texts.
This paper separates binary relations from the kind of relations
that are needed in relational databases.
Its treatment of functions supports
both computer science in general
and the kind of relations needed in databases.
As a sample application
this paper shows how the mathematical theory of relations
naturally leads to the relational data model
and how the operations on relations
are by themselves already a powerful vehicle for queries.
\end{abstract}

\newcommand{\cpp}{\hbox{{\tt C++}}}

\newtheorem{theorem}{Theorem}{}
\newtheorem{definition}{Definition}{}
\newtheorem{example}{Example}{}

\newcommand{\Nat}{\ensuremath{\mathcal{N}}} 
\newcommand{\Int}{\ensuremath{\mathcal{I}}} 
\newcommand{\Rat}{\ensuremath{\mathcal{Q}}} 
\newcommand{\Rea}{\ensuremath{\mathcal{R}}} 
\newcommand{\ExtRe}{\ensuremath{\mathcal{R}^{++}}} 
\newcommand{\Flpt}{\ensuremath{\mathcal{F}}} 

\newcommand{\Alp}{\ensuremath{\mbox{\textbf{A}}}} 
\newcommand{\AlpStar} {\ensuremath{\Alp ^\ast}}

\newcommand{\A}{\ensuremath{\mathcal{A}}} 
\newcommand{\Pwst}{\ensuremath{\mathcal{P}}} 
\newcommand{\T}{\ensuremath{\mathcal{T}}} 
\newcommand{\cart}{\ensuremath{\mbox{\textsc{cart}}}} 
\newcommand{\apl}{\ensuremath{\mbox{\textsc{apl}}}}

\newcommand{\pair}[2]{\ensuremath{\langle #1,#2 \rangle}}
\newcommand{\triple}[3]{\ensuremath{\langle #1,#2,#3 \rangle}}
\newcommand{\vc}[2]{\ensuremath{#1_0,\ldots,#1_{#2-1}}}

\newcommand{\id}{\ensuremath{\mbox{id}}}
\newcommand{\argt}{\ensuremath{\mbox{argt}}}
\newcommand{\prd}{\ensuremath{\mbox{prod}}}
\newcommand{\sq}{\ensuremath{\mbox{sq}}}
\newcommand{\tr}{\ensuremath{\triangleright}}

\newcommand{\iot}[1]{\ensuremath{#1}}

\newcommand{\cities}{\ensuremath{ \mbox{ \textsc{cities}} }}
\newcommand{\parts}{\ensuremath{ \mbox{ \textsc{parts}} }}
\newcommand{\projects}{\ensuremath{ \mbox{ \textsc{projects}} }}

\newcommand{\sid}{\ensuremath{ \mbox{ \textsc{sid}} }}
\newcommand{\sname}{\ensuremath{ \mbox{ \textsc{sname}} }}
\newcommand{\city}{\ensuremath{ \mbox{ \textsc{city}} }}
\newcommand{\pid}{\ensuremath{ \mbox{ \textsc{pid}} }}
\newcommand{\pname}{\ensuremath{ \mbox{ \textsc{pname}} }}
\newcommand{\pqty}{\ensuremath{ \mbox{ \textsc{pqty}} }}
\newcommand{\LEQ}{\ensuremath{ \mbox{ \textsc{leq}} }}

\newcommand{\suppliers}{\ensuremath{ \mbox{ \textsc{suppliers}} }}

\newcommand{\lee}{\ensuremath{ \mbox{ \textsc{lee}} }}
\newcommand{\tulsa}{\ensuremath{ \mbox{ \textsc{tulsa}} }}
\newcommand{\poe}{\ensuremath{ \mbox{ \textsc{poe}} }}
\newcommand{\ray}{\ensuremath{ \mbox{ \textsc{ray}} }}
\newcommand{\taos}{\ensuremath{ \mbox{ \textsc{taos}} }}
\newcommand{\hose}{\ensuremath{ \mbox{ \textsc{hose}} }}
\newcommand{\tube}{\ensuremath{ \mbox{ \textsc{tube}} }}
\newcommand{\shim}{\ensuremath{ \mbox{ \textsc{shim}} }}

\newcommand{\rid}{\ensuremath{ \mbox{ \textsc{rid}} }}
\newcommand{\rqty}{\ensuremath{ \mbox{ \textsc{rqty}} }}
\newcommand{\where}{\ensuremath{ \mbox{ \textsc{where}} }}
\newcommand{\pc}{\ensuremath{ \textsc{pc} }}

\newcommand{\parent}{\ensuremath{ \textsc{parent} }}
\newcommand{\child}{\ensuremath{ \textsc{child} }}
\newcommand{\mary}{\ensuremath{ \textsc{mary} }}
\newcommand{\john}{\ensuremath{ \textsc{john} }}
\newcommand{\alan}{\ensuremath{ \textsc{alan} }}
\newcommand{\joan}{\ensuremath{ \textsc{joan} }}

\newcommand{\answer}{\ensuremath{ \textsc{answer} }}

\section{Introduction}

Mathematics is more useful than computer scientists tend to think.
I am not talking about specialized areas of computer science
with strongly developed mathematical theories,
like graph theory in computer networks or
posets and lattices for programming language semantics.
These are solid extensions
of the traditional notion of Applied Mathematics.

No, I'm talking about the mundane mathematics
that falls into the cracks between these well-developed areas:
basic facts about sets, functions, and relations.
I refer to the material that gets relegated to the desultory miscellanies
added to texts under the name of ``set-theoretic preliminaries''.
The reason for the perfunctory way
in which these are compiled
is that computer scientists are ambivalent about basic set theory:
they don't care for that stuff, yet dare not omit it.

My own lack of understanding in this area resulted in difficulties
with the relational model.
Accordingly, I include it as a case study of how basic set theory could
have clarified these difficulties, which are also found in the literature.

The ambivalence concerning the set-theoretic preliminaries
is understandable.
Is a proper treatment not
going to lead to an unacceptably large mass of definitions and theorems?
After all, set theory is nothing
if it is not done rigorously.
In other parts of mathematics
you can start right away with what interests you
and you refer to something else for the foundation.
No such help is available
when you deal with the foundation itself.

This paper is intended to improve this situation.
My model is Halmos's Na\"{\i}ve Set Theory \cite{hlm60},
which shows that one can be precise enough
and yet easily digestible without losing anything essential.
Halmos's book was aimed at normal mathematicians;
that is, those who have no interest in set theory,
yet cannot do without it.
It is therefore ideal for computer scientists,
who are in the same position.
Though set theory is no different for mathematicians
than it is for computer scientists,
it does make sense to rewrite the first part of \cite{hlm60}
into a sort of Halmos for computer scientists.
This is what I am trying to do here.

Halmos had to maintain a delicate balance
between conflicting requirements.
On the one hand he wanted to avoid an arid listing
of the facts and definitions that normal mathematicians need.
Such a bare listing would not do justice
to the intrinsic interest of the subject.
On the other hand, he did not want to get sucked
into the depth, and richness, of the subject.
I have tried to follow his example.
On the one hand, I give more than a minimal listing of facts.
For example, I even start with a \emph{history} of set theory.
To maintain the balance, I kept it to less than five hundred words.
In the sequel I have tried to continue this balance.

These considerations have resulted in a paper
that is difficult to classify.
It is part review, part tutorial.
As a result of approaching certain subjects
important for computer scientists, like relations,
in the way that mathematicians had learned to do
by the middle of the 20th century,
certain new and useful results come out,
so that this paper also contains recent research.
But most of all, the purpose is methodological:
to show that, before breaking new ground,
we should go back to the basic abstract mathematics
that became the consensus of mathematicians in the 1930s
and was codified by Bourbaki in the
\emph{Fascicule de R\'esultats} \cite{brbk39} of their
\emph{Th\'eorie des Ensembles} \cite{brbk54}.

\section{Sets}

The development of the calculus in the 18th century
was a spectacular practical success.
At the same time, philosophers poked fun
at the mathematicians for the peculiar logic
used in reasoning about infinitesimals
(sometimes they were treated as zero; sometimes not).
In the 19th century, starting with Cauchy,
analysis developed as the theory underlying the calculus.
The goal was to do away with infinitesimals and explain
the processes of analysis in terms
of limits, rationals, and, ultimately, integers
as the bedrock foundation\footnote{
Leopold Kronecker (1823-1891) is widely quoted as having uttered:
``Die ganze Zahl schuf der liebe Gott, alles Uebrige ist Menschenwerk.''
(God made the integers, all else is the work of man).
}.

In the late 19th century Cantor
was led to create set theory
in response to further problems in analysis.
This theory promised to be an even deeper layer of bedrock
in terms of which even the integers,
including hierarchies of infinities, could be explained\footnote{
Professor Kronecker did not take kindly to these new developments
and intervened to thwart Cantor's career advancement.
}.

For some decades set theory
was an esoteric and controversial theory,
plagued by paradoxes.
The axiomatic treatment of Zermelo and Fraenkel
made set theory into respectable mathematics.
A sure sign of its newly achieved status
was that Bourbaki started their great multivolume treatise
on analysis with a summary of set theory \cite{brbk54}
to serve as foundation of the entire edifice.

When set theory is not just viewed
as one of the branches of mathematics,
but as the foundation of all mathematics,
it is tempting to conclude
that the proper, systematic way to teach
any kind of mathematics is to start with set theory.
Combine this idea with the curious notion
that children are in school not just to learn to reckon,
but to do ``Math'', and you get a movement
like ``New Math'' according to which
children are taught sets before they get to counting.

By the time the disastrous results
of this approach were recognized,
set theory was out of favour.
Even pure mathematicians were no longer impressed
by the edifice erected by Bourbaki.
Set theory was relegated to the status
of one of the many branches of mathematics;
one distinguished by the utter lack of applications.
Mathematicians interested in foundations of mathematics
and theoretically minded computer scientists
use categories rather than set theory.
A characteristic quote from this part of the world is:
``Set theory is the biggest mistake in mathematics
since Roman numerals.''


For the kind of things that these people do,
category theory may well be superior.
What I hope to show here
is that a bit of basic old-fashioned set theory
goes a long way in improving what the rest of us do. 

\subsection{Set membership and set inclusion}

What is a set?

I shall remain evasive on this point.
In geometry, there are definitions of things
like triangles and parallelograms,
but not of points and lines.
The latter are undefined primitive objects
in terms of which other geometrical figures are defined.
All we are allowed to assume about points and lines
is what the axioms say about the relations between them
(like there being one and only one line on which two given points lie).

In the same way it is not appropriate to ask ``What is a set?''.
All we know about sets is
that they enter into a certain relation
with another set or with an element
and that these relations
have certain properties.
Some of these I review here without attempting completeness.

If $S$ is a set, then $x \in S$
means that $x$ is an \emph{element}
(or \emph{member}) of $S$.
The symbol $\in$ stands for the \emph{membership} relation.
If $S$ and $T$ are sets, then $S \subset T$ means
that every element of $S$ (if any) is an element of $T$ (if any).
We say that $S$ is a \emph{subset} of $T$
and that $T$ is a \emph{superset} of $S$.
Thus for every set $S$, we have $S \subset S$.
Many authors use ``$\subseteq$'' to indicate the subset-superset relation.

If we have both $S \subset T$ and $T \subset S$,
then $S$ and $T$ have the same elements
and we write $S = T$.
An important property of sets is that
if $S$ and $T$ have the same elements,
then $S$ is the same set as $T$.
That is, a set is completely determined by the elements it contains.
A set is not one of those wholes
that are more than the sum of their parts.

To specify a set we only need to tell what elements it has.
We can do this in two ways:
element by element or by giving a rule for membership.
The notation for the first way is to list the elements between braces, as in
\begin{eqnarray*}
S = \{\spadesuit,\mbox{p},\dagger,5,\P\},
\end{eqnarray*}
which says that the elements of $S$ are
$
\spadesuit
$,
$
\mbox{p}
$,
$
\dagger
$,
$
5
$, and
$
\P
$, and that there are no other elements.
As a set is determined by its elements,
the order in the listing is immaterial.

The other specification method is by a rule
that determines the elements of a set; for example
\begin{eqnarray*}
\{x \mid x \in \Nat
         \mbox{ and } x \mbox{ is not divisible by } 2
\}    \nonumber
\end{eqnarray*}
for the set of odd numbers, where \Nat\ is the set of natural numbers.
In this style $\{\spadesuit,\mbox{p},\dagger,5,\P\}$
can be specified as
$$
\{x \mid
x = \spadesuit \mbox{ or } 
x = \mbox{p}   \mbox{ or } 
x = \dagger    \mbox{ or } 
x = 5    \mbox{ or } 
x = \P
\}.
$$

The set with no elements is called
the \emph{empty} set or the \emph{null} set,
which we write as $\{\}$ or as $\emptyset$.
The empty set is a subset of every set\footnote{
It is even a subset of the empty set.
}.

%

\paragraph{Set operations}

If $S$ and $T$ are sets, then
\begin{eqnarray*}
S \cup T &\stackrel{\mathrm{def}}{=}& \{x \mid x \in S \mbox{ or } x \in T\}     \\
S \cap T &\stackrel{\mathrm{def}}{=}& \{x \mid x \in S \mbox{ and } x \in T\}    \\
S \setminus T &\stackrel{\mathrm{def}}{=}& \{x \mid x \in S \mbox{ and } x \not\in T\}
\end{eqnarray*}
are also sets.
$S \cup T$ is known as the \emph{union} of $S$ and $T$,
$S \cap T$ as the \emph{intersection},
and $S \setminus T$ as the \emph{set difference}.
The expression $x \not\in T$ occurring in the definition
of $S \setminus T$ means that $x$ is not a member of $T$. 

One can characterize 
$S \cup T$ and $S \cap T$, respectively,
as the least common superset and greatest common subset
of $S$ and $T$.

\paragraph{Sets of sets}
Elements of a set can themselves be sets.
For every set $S$ we define the \emph{powerset}
$\Pwst(S)$ of $S$ as the set of all subsets of $S$.
For example,
\begin{eqnarray*}
\Pwst(\{a,b,c\}) &=&
\{
\{\},
\{a\},
\{b\},
\{c\},
\{a,b\},
\{a,c\},
\{b,c\},
\{a,b,c\}
\}                                      \\
\Pwst(\{\}) &=&
\{\{\}\}                                \\
\Pwst(\{\P\}) &=&
\{\{\}, \{\P\}\}
\end{eqnarray*}

Let $S$ be a nonempty set of sets.
Then
$\cup S$ is defined as $\{x \mid \exists S' \in S\,.\,  x \in S' \}$
and
$\cap S$ as $\{x \mid \forall S' \in S\,.\,  x \in S' \}$.

Let $S$ be a set of sets
and let $S_0$ and $S_1$ be non-empty subsets of $S$.
Then
$S_0 \subset S_1$ implies $\cup S_0 \subset \cup S_1$.
It also implies that $\cap S_1 \subset \cap S_0$.

\paragraph{The na\"{\i}ve point of view}
What makes sets interesting is that their elements can be sets.
Then of course one also has sets
that contain sets that contain sets, and so on.
It is easy to get carried away
and consider things like $\{x \mid x \not\in x\}$.
To assume that such a thing is a set leads to contradiction.
As a result, mathematicians have learned to be careful.
This care takes the form of axioms that explicitly state what
things are allowed to be sets. Nothing is a set unless these axioms
say it is. This is \emph{axiomatic} set theory.

In this paper I do not try to justify
the existence of the sets that we want to talk about.
That is the ``na\"{\i}ve point of view''.
According to axiomatic set theory,
no things exist except sets.
It is na\"{\i}ve to assume that a set
such as $\{\spadesuit,\mbox{p},\dagger,5,\P\}$ exists.
Axiomatic set theory requires one to justify
the existence \emph{as sets} of
$\spadesuit$,
$\mbox{p}$,
$\dagger$,
$5$, and
$\P$.

It is the hallmark of na\"{\i}vet\'e
to emphasize, as I often will do,
that the elements of a set are sets:
in axiomatic set theory,
every nonempty set answers this description.

As an example of familiar objects being reconstructed as sets,
let us consider numbers.
One of the things set theory is expected to do
is to explain and to justify the various types of numbers.
When starting with sets only,
we cannot say that
$\{\spadesuit,\mbox{p},\dagger,5,\P\}$
has five elements,
or even that this set has more elements than
$\{\spadesuit,\mbox{p},5,\P\}$.
``Five'' and ``more'' needs to be explained in terms of sets.
We cannot even say that (or whether) a set has a finite
number of elements without a set-theoretic explanation of finiteness.

While remaining na\"{\i}ve,
we should still get \emph{some} idea
of how axiomatic set theory could justify
the existence of the kind of sets that we work with.
Consider the natural numbers 0, 1, 2, $\ldots$
Strictly speaking such numbers are either ``ordinal''
or ``cardinal''.
Ordinal numbers refer to locations in a linear order;
cardinal numbers refer to sizes of sets.
The construction of ordinal numbers in terms of sets is simplest,
so we'll just consider that.
As the natural numbers are all finite anyway, the distinction
between cardinal and ordinal numbers need not concern us.

The axioms guarantee for every set $S$
the existence of its unique successor set $S^+$,
which is $S \cup \{S\}$.
This at least permits us to say that $S^+$ has
\emph{one more} element than $S$,
even if we don't know how to count the elements of $S$.

As the axioms also allow
the existence of the empty set $\emptyset$,
we have also the sets  $\emptyset^+$, $\emptyset^{++}$, and so on,
all the way up to infinity (and beyond).
These are considered
to be the ordinal numbers 0, 1,  2, and so on, respectively.
Before set theory it was already shown
how one could get from the set \Nat\ of natural numbers
successively the set \Int\ of integers, the set \Rat\ of rationals (that's simple),
and, from the rationals, the set \Rea\ of reals (not so simple).

If $S^+=S\cup\{S\}$, then $S \subset S^+$.
As a result,
$
\emptyset \subset
\emptyset^+ \subset
\emptyset^{++} \subset
\cdots
$
In fact, one can see that
$\emptyset =\{\}$,
$\emptyset^+ =\{\emptyset\}$,
$\emptyset^{++} =\{\emptyset,\emptyset^+\}$,
$\ldots$ :
Every ordinal number is the set of its predecessors.
So we could just consider the natural numbers as finite ordinals
and consider the natural number $0$ to be $\emptyset$
and every positive natural number $n$ to be $\{0,\ldots,n-1\}$.
Instead of ``for all $i = 0, \ldots, n-1$''
we could be rid of the ambiguous ``$\ldots$''
and write ``for all $i \in n$''.
Though this consequence of a natural number $n$ being a set
is not widely used,
it is concise and unambiguous.

%

\subsection{Partitions}

Let $P$ be a set of non-empty subsets of a set $S$.
Then we have $\cup P \subset S$.
If we also have $\cup P = S$, then $P$ is a \emph{cover} for $S$.
The elements of $P$ are called the \emph{cells} of the cover.
If a cover $P$ for $S$ is such that the sets of $P$ are mutually
disjoint, then $P$ is a \emph{partition} of $S$.

A partition $P'$ of $S$ is said to be \emph{finer} than $P$
if, for every cell $C'$ of $P'$,
there is a cell $C$ of $P$ such that $C' \subset C$.
This makes every partition finer than itself,
which is fine (in mathematics).
It makes $\{\{x\} \mid x \in S \} $ the finest partition of $S$,
and $\{\{S\}\}$ the coarsest.

\subsection{Pairs}

The order in which we list the elements of a $\{a,b\}$ is
an artifact of the notation.
The equality of $\{a,b\}$ and $\{b,a\}$
can perhaps be seen most clearly
by writing both as
$
\{x \mid x = a \vee x = b \}.
$

How does one use sets not only to bundle $a$ and $b$ together
but also to place them in a certain order?
The usual way to do this is based on the observation that,
though 
$ \{a,b\} $
and
$ \{b,a\} $
are the same set,
$ \{\{a\},\{a,b\}\} $
and
$ \{\{b\},\{a,b\}\} $
do not have the same elements, so are not the same set.
We call 
$ \{\{a\},\{a,b\}\} $
a \emph{pair}, and write $\langle a, b \rangle$.
The \emph{left element} of $\pair{a}{b}$ is $a$;
its \emph{right element} is $b$.

In a similar way one can define triples
$\langle a, b, c \rangle$
as 
$\langle \langle a, b \rangle, c \rangle$
or
$\langle a, \langle b, c \rangle \rangle$
quadruples
$\langle a, b, c, d \rangle$,
and so on.

\section{Binary relations}
\label{binRelSec}

A relation among sets associates some of the elements
in each of the sets with each other.
In this section, I present only binary relations:
relations between two sets or between a set and itself.
Relations between any number of sets are treated later,
independently of binary relations.

Everything so far in this paper is standard.
However, from now on
authors differ on some of the terminology and definitions,
though the underlying concepts are universally accepted.
Hence the occasional Definition from this point onward.

\begin{definition}[binary relation, source, target, extent, $\leftrightarrow$]
A \emph{binary relation} is a triple of the form
$\triple{S}{T}{E}$,
where $S$ and $T$ are sets
and $E$ is a set of pairs.
If $E$ is nonempty,
then $\pair{x}{y} \in E$ implies $x \in S$ and $y \in T$.
$S$ is the \emph{source}, $T$ is the \emph{target},
and $E$ is the \emph{extent}
of $\triple{S}{T}{E}.$
The set of all binary relations
with source $S$ and target $T$
is denoted by the expression $S \leftrightarrow T$.
\end{definition}

\begin{definition}[binary Cartesian product]
\label{binCart}
The \emph{binary Cartesian product} of sets
$S$ and $T$ is defined to be
$\{\pair{x}{y}\mid x \in S \wedge y \in T \}$.
It is written as $S \times T$.
\end{definition}

\begin{definition}[empty binary relation, universal binary relation]
The \emph{empty binary relation} in $S \leftrightarrow T$
is the one of which the extent is the empty set of pairs.
The  binary relation in $S \leftrightarrow T$ that has
$S \times T$ as extent is the \emph{universal binary relation},
denoted $U(S,T)$.
\end{definition}

\subsection{Set-like operations and comparisons}
\label{setLikeOp}
Among the binary relations in $S \leftrightarrow T$
certain operations are defined
that mirror set operations:
\begin{eqnarray*}
\triple{S}{T}{E_0} \cup \triple{S}{T}{E_1} & \stackrel{def}{=} &
              \triple{S}{T}{E_0\cup E_1}         \\
\triple{S}{T}{E_0} \cap \triple{S}{T}{E_1} & \stackrel{def}{=} &
              \triple{S}{T}{E_0\cap E_1}         \\
\triple{S}{T}{E_0} \setminus \triple{S}{T}{E_1} & \stackrel{def}{=} &
              \triple{S}{T}{E_0\setminus E_1}
\end{eqnarray*}
Comparison of sets carries over in a similar way to binary relations:
\begin{eqnarray*}
\triple{S}{T}{E_0} \subset \triple{S}{T}{E_1} & \mbox{iff} & E_0 \subset E_1 \\
\end{eqnarray*}

\subsection{Properties of binary relations}
Consider a binary relation $\triple{S}{T}{E}$.
It has any subset of the following properties:
\begin{itemize}
\item
{\bf total:}
for every $x \in S$ there exists $y \in T$
such that $\pair{x}{y} \in E$.
\item
{\bf single-valued:}
for all $x \in S$,
$y_0 \in T$, and
$y_1 \in T$,
$\pair{x}{y_0} \in E$ and
$\pair{x}{y_1} \in E$ imply
$y_0 = y_1$.
\item
{\bf surjective:}
for every $y \in T$
there exists $x \in S$ such that $\pair{x}{y} \in E$.
\item
{\bf injective:}
for all $y \in T$,
$x_0 \in S$, and
$x_1 \in S$,
$\pair{x_0}{y} \in E$ and
$\pair{x_1}{y} \in E$ imply
$x_0 = x_1$.
\end{itemize}

\begin{definition}
\label{funcBinRel}
A \emph{partial function} is defined to be
a single-valued binary relation.
A \emph{functional binary relation} is
a partial function that is total.
\end{definition}

We reserve ``function'' for the similar concept that
will be defined later independently of binary relations.
Some authors, however, prefer to regard a function
as the special case of a binary relation
that is single-valued and total.

\subsection{Other operations on binary relations}

In addition to the set-like operations defined in section~\ref{setLikeOp},
the following are defined for binary relations.

The \emph{inverse}
of the binary relation
$\triple{S}{T}{E}$
is given by
$$\triple{S}{T}{E}^{-1} =
\triple{T}{S}{\{\pair{y}{x}\mid \pair{x}{y} \in E\}}.
$$
For every binary relation $r$ we have that $(r^{-1})^{-1} = r.$

It is apparent from this definition that
every binary relation has an inverse:
no properties are needed.
For example, every partial function $f$ has an inverse $f^{-1}$,
\emph{when regarded as binary relation}.
But $f^{-1}$ is not necessarily a partial function:
for that to be true, $f$ needs to be injective.

\begin{definition}[composition]
The \emph{composition} ``;'' of two binary relations is defined
when the target of the first is the source of the second.
In that case it is defined by
$$
\triple{S}{T}{E} ; \triple{T}{U}{F}
\stackrel{def}{=}
\triple{S}{U}{
\{ \pair{x}{z} \mid
\exists y \in T. \pair{x}{y} \in E \wedge \pair{y}{z} \in F 
\}
}
$$
\end{definition}

If $r_0;r_1$ is defined,
then so is $r_1^{-1};r_0^{-1}$,
and it equals $(r_0;r_1)^{-1}$.

\subsection{Endo-relations}

A binary relation where source and target are the same set
may well be called an \emph{endo-relation}.
Some of the most commonly used binary relations are of this kind.

For every set $S$ we define the \emph{identity} on $S$
as $\id_S \stackrel{def}{=} \triple{S}{S}{\{\pair{x}{x} \mid x \in S\}}$.

The identity relations allow, in combination with composition,
succinct characterizations of
various properties that a relation $r$ in $S \leftrightarrow T$
may have:
\begin{eqnarray*}
\mbox{single-valued} & \mbox{iff} & r^{-1};r \subset \id_T      \\
\mbox{surjective} & \mbox{iff} & r^{-1};r \supset \id_T      \\
\mbox{injective} & \mbox{iff} & r;r^{-1} \subset \id_S      \\
\mbox{total} & \mbox{iff} & r;r^{-1} \supset \id_S
\end{eqnarray*}

\paragraph{Equivalence relations}
Let us consider the endo-relation $r = \triple{S}{S}{E}$.
\begin{center}
\begin{tabular}{rl}
If $\id_S \subset r$,&then $r$ is said to be \emph{reflexive}.  \\
If $r = r^{-1}$,&then $r$ is said to be \emph{symmetric}.  \\
If $r;r \subset r$,&then $r$ is said to be \emph{transitive}.
\end{tabular}
\end{center}
An endo-relation that has these three properties is called an
\emph{equivalence}.

Consider
$\{\{y \mid \pair{x}{y} \in E\} \mid x \in S\}$.
If $r$ is reflexive, then this is a cover for $S$.
If $r$ is an equivalence, then it is a partition.

\paragraph{Some order relations}
If we drop from equivalence the symmetry requirement,
then we are left with
a weak type of order called a \emph{pre-order}.
Even the inclusion relation
among the subsets of a set is stronger:
in addition to reflexivity and transitivity,
it has the property of also being \emph{antisymmetric},
which can be defined as $r \cap r^{-1} \subset id_S$. 
An antisymmetric pre-order is a \emph{partial order}.

Partial orders are called thus because not every pair of elements
has to be in the relation.
A partial order $r$ such that $r \cup r^{-1} = U(S,S)$,
the universal relation on $S$,
is said to be \emph{order-total}.
The qualification ``order-'' serves to avoid confusion
with the property of being total that was defined earlier.
A partial order that is also order-total
is called a \emph{total order}, or a \emph{linear order}.

\section{Functions}

The following definition is independent of the definition of the
functional binary relation in Definition~\ref{funcBinRel}.

\begin{definition}[function, source, target, map, $\rightarrow$]
\label{funDef}
A \emph{function} consists of a set $S$, its \emph{source},
a set $T$, its \emph{target}, and a \emph{map},
which associates to every element of $S$ a unique element of $T$.
We write $S \rightarrow T$ for the set
of all functions with $S$ as source and $T$ as target.
\end{definition}

$S$ is often called the ``domain'' of the function.
However, I prefer to reserve ``domain''
for the different meanings that are entrenched
in subspecializations of computer science
such as programming language semantics, constraint processing, and databases.
The arrow notation suggests ``source''
as the needed alternative for ``domain'';
``target'' is its natural counterpart.

The set $S \rightarrow T$ is said to be
the \emph{type} of $f \in S \rightarrow T$.
One often sees ``$f:S\rightarrow T$''
when $f \in S\rightarrow T$ is meant.

$S$ and $T$ are typically nonempty, but need not be.
$S$ and $T$ may or may not be the same set.

If the map of $f$ associates the element $y$ in $T$
with the element $x$ in $S$,
then we call $y$ the \emph{value} of the function
\emph{at} the \emph{argument} $x$.
We may also say that $f$ \emph{maps} $x$ \emph{to} $y$.
We write $y$ as $f(x)$ to emphasize
that $y$ is determined the argument $x$.

\begin{example}
Given $f \in S \rightarrow T$,
we can define the binary relation
$\triple{S}{T}{\{\pair{x}{f(x)}\mid\ x \in S\}}$.
It is a functional binary relation.

Given a functional binary relation $\triple{S}{T}{E}$,
we can define a function in $S\rightarrow T$
whose map associates $x$ with the unique $y$ such that
$\pair{x}{y} \in E$, for every $x \in S$.  
\end{example}


The arrow notation $S \rightarrow T$ makes it easy to remember
which set is the source and which is the target.
$S \rightarrow T$ is sometimes written as $T^S$.
This notation makes it easier to confuse source and target.
Here is a way to remember which is which.
Let us define, for some sets at least,
$|S|$ as the number of elements of $S$.
Then it can be seen that $|S \rightarrow T| = |T|^{|S|}$.

This formula is remarkably resistant to perverse special cases
of which we now consider a few.
Consider $\{x\} \rightarrow T$,
where the source is a one-element set.
As any such function has to associate one element of $T$
with $x$, $T$ cannot be empty.
It is easy to count the functions of this type:
there are as many such functions as there are elements in $T$. 
The set $\emptyset \rightarrow T$ is reckoned to have one
function in it: there is only one way of associating an element
of the target with each element of the source if there are no elements
in the source.

\subsection{Multiple arguments}
Functions have one argument.
But source and target can themselves have structure
and this can be arranged in such a way as to suggest multiple arguments.

Consider $f \in S_0 \rightarrow (S_1 \rightarrow T)$.
With $f$ of this type,
$x_0 \in S_0$ implies that $f(x_0) \in S_1 \rightarrow T$
and
$x_1 \in S_1$ implies that $f(x_0)$, if given argument $x_1$,
yields as value $(f(x_0))(x_1) \in T$.

Instead of $(f(x_0))(x_1)$ we will write $f(x_0,x_1)$.
Similarly, we will regard $f(x_0, \ldots, x_{n-1})$
as shorthand for the value of a function in
$$
S_0 \rightarrow
  (S_1 \rightarrow \cdots \rightarrow (S_{n-1} \rightarrow T) \cdots)
$$
at arguments
$x_0 \in S_0, \ldots, x_{n-1} \in S_{n-1}$.
This construction is called ``Currying'' after the logician
Haskell B. Curry.
\begin{example}
Consider $g(x) = y$ with $g \in U \rightarrow V$.
Here $y$ is determined by two things: $g$ and $x$.
Hence there is a function, say $\alpha$,
that has value $y$ for arguments $g$ and $x$.
This function is in $(U \rightarrow V) \rightarrow (U \rightarrow V)$.
Compare this with two-argument Currying as in 
$f \in S_0 \rightarrow (S_1 \rightarrow T)$
and let $S_0$ correspond to $(U \rightarrow V)$ and
$S_1$ to $U$.
Then we see that $\alpha$ is a two argument function with $g \in U \rightarrow V$
as first argument and $x \in U$ as second argument.
In functional programming, this is the function called \textsc{apply}.
\end{example}

\subsection{Expressions}
Definition~\ref{funDef} specifies a map as being part of a function.
It does not specify how this map
associates source elements with target elements.
One of the ways in which this can be done
is by an expression $E$
typically containing $x$
if the value of $E$ is defined for every value of $a \in S$
substituted for $x$ and if the values of $E$ are in $T$.
We can then find $y$ as the result of evaluating $E$.

The expression can be a program or it can be
an arithmetic or symbolic expression.
To indicate that the map associates $a$ with $E[x/a]$,
one writes $x \mapsto E$.

$E$ by itself does not specify a function,
as such a specification requires the source and the target.
$E$ may not even be sufficient as specification of the map.
For example, when $E$ is $x^2+y$, then
$x \mapsto E$ and $y \mapsto E$
are the maps of different functions.
If there is an expression that specifies the map of a function,
then there are typically other ones as well.
For example, if $x \in \Rea$, then one may prefer
$(x+1)^2$ to $x^2+2x+1$.
A reason could be that one prefers to avoid
multiple occurrences of a variable.

The notation $x \mapsto E$ is similar $\lambda x.E$,
as in lambda calculus.
The arrow notation has the advantage of requiring one symbol
less than $\lambda x.E$. Moreover, when there are two
arguments, as we have here with $x$ and $E$, infix
is the most convenient notation
(compare ``$2+2$'' with ``$+2.2$'').

\begin{example}
$f \in \Nat \rightarrow \Nat$ with map
$n \mapsto ((1+\surd 5)^n - (1-\surd 5)^n)/(2^n \surd 5$)
is the function that gives the $n$-th Fibonacci number.
\end{example}

\subsection{Properties of functions}

Note the asymmetry in the definition of
a function $f \in S\rightarrow T$:
with \emph{every} $x \in S$ the map of the function
associates a \emph{unique} $y \in T$.

The function $f \in S \rightarrow T$ may associate
more than one $x \in S$ with the same $y \in T$.
If, on the other hand, $f(x_1) = f(x_2)$ implies
that $x_1 = x_2$, then this is a property that not all functions have.
A function that has this property is said to be \emph{injective}.

It is not necessarily the case that every $y \in T$ is equal
to $f(x)$ for some $x \in S$.
If $f$ has this additional property,
then it is said to be \emph{surjective}. 

To be able to count elements in a set
we have to define a suitable notion of number.
It would be nice and tidy if this notion
of number of elements in a set were based on sets.
As we have not done this,
we cannot officially count the elements in a set.
But with injectivity and surjectivity
we can at least compare sizes of sets:
if $f \in S \rightarrow T$ is surjective, then
there have to be at least as many elements
in $S$ as there are in $T$;
if $f \in S \rightarrow T$ is injective,
then there have to be at least as many elements
in $T$ as there are in $S$.
If $f$ is both injective and surjective
(such a function is said to be \emph{bijective}),
then $S$ and $T$ have the same number of elements,
however ``number'' is defined.

If $f \in S \rightarrow T$ is a bijection,
then there exists a function $g$ in $T \rightarrow S$
that associates with every $y \in T$ the unique $x \in S$
such that $f(x) = y$.
This $g$ is uniquely determined by $f$,
is written $f^{-1}$,
and is called the \emph{inverse} of $f$.
It is a bijection; hence has an inverse $(f^{-1})^{-1}$,
which is equal to $f$.

If there exists a bijection in $\iot{n} \rightarrow S$,
for some natural number $n$,
then we can define $S$ to have $n$ elements.
In that case, $S$ is said to be \emph{finite},
otherwise \emph{infinite}.
$S$ is \emph{countably} infinite if there exists a bijection in $\Nat \rightarrow S$.
If two sets are infinite,
then it is not necessarily the case
that there is a bijection between them.
For example, there is no bijection in $\Rat \rightarrow \Rea$,
where \Rat\  and \Rea\ are the sets of rational and real
numbers, respectively.

\begin{definition}[identity, restriction, insertion]
The function in $S \rightarrow S$ with map $x \mapsto x$
is $\id_S$, the \emph{identity function on} $S$.
(I count on the context to prevent confusion with the identity
that is a binary relation.
)
\\
Let $f$ be a function in $S\rightarrow T$ and let $S'$ be a set.
The \emph{restriction} of $f$ to $S'$ is written
$f \downarrow S'$ and is defined\footnote{
Some authors restrict the notion of restriction to the case
where $S' \subset S$.
}
as the function in 
$S \cap S' \rightarrow T$ that has the mapping
that associates
$f(x) \in T$ with every $x$ (if any) in $S \cap S'$.
\\
If $S' \subset S$, then
$\id_S \downarrow S'$ is called the \emph{insertion function}
determined by the subset $S'$ of $S$.
\end{definition}
Another function solely determined by subsets
is the \emph{characteristic function} for $S'$,
which is in $S \rightarrow 2$
and maps $x$ in $S$ to $1$ if $x \in S'$
and maps $x$ to $0$ otherwise.

\subsection{Function sum}

\begin{definition}[function sum]
Functions
$f_0 \in S_0 \rightarrow T_0$
and
$f_1 \in S_1 \rightarrow T_1$
are \emph{summable} iff for all $x \in S_0 \cap S_1$, if any,
we have $f_0(x) = f_1(x)$.
If this is the case, then $f_0+f_1$, the \emph{sum}
of $f_0$ and $f_1$, is the function in $(S_0 \cup S_1) \rightarrow (T_0 \cup T_1)$
with map
$x \mapsto f_0(x)$ if $x \in S_0$
and
$x \mapsto f_1(x)$ if $x \in S_1$.
\end{definition}

\begin{example}
Let
$f_0 \in \{a,b\}\rightarrow\{0,1\}$
such that $a \mapsto 0$ and $b \mapsto 1$.
Let
$f_1 \in \{b,c\}\rightarrow\{0,1\}$
such that $b \mapsto 1$ and $c \mapsto 0$.
Then $f_0$ and $f_1$ are summable and
$(f_0+f_1) \in \{a,b,c\}\rightarrow\{0,1\}$
such that $a \mapsto 0$,
$b \mapsto 1$, and
$c \mapsto 0$.
\end{example}

\emph{Remarks about summability:}
Given
$f_0 \in S_0 \rightarrow T_0$
and
$f_1 \in S_1 \rightarrow T_1$.
$f_0 \downarrow S_1$ and $f_0$ are summable,
as are 
$f_0 \downarrow S_1$ and $f_0\downarrow (S_0\setminus S_1)$
and
$f_0
= f_0 + (f_0 \downarrow S_1)
= (f_0\downarrow S_1) + (f_0 \downarrow (S_0\setminus S_1)
$.

If $f_0$ and $f_1$ are summable, then
\begin{itemize}
\item
$f_0 +f_1
= f_0 + f_1\downarrow (S_1\setminus S_0)
= f_1 + f_0\downarrow (S_0\setminus S_1)
$
\item
if, moreover, $S_0 \cap S_1$ is nonempty,
then $T_0 \cap T_1$ is.
\end{itemize}

\begin{figure}[htbp]
\begin{center}
\epsfig{file=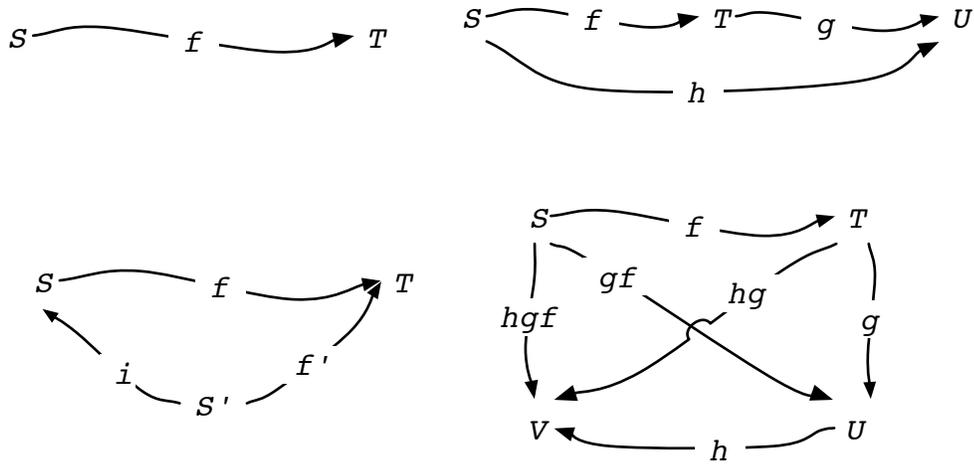}
\end{center}
\caption{\label{fig:fourExamples}
Top left: $f \in S \rightarrow T$.
Top right: $h = g \circ f$.
Bottom left:  $S' \subset S$ and $i \in S' \rightarrow S$ is the insertion
function of $S'$ as a subset. The diagram shows
$f' = f \downarrow S' = f \circ i$. 
Bottom right: $gf = g\circ f$, $hg = h\circ g$, and
$hgf= h \circ (g \circ f) = (h\circ g) \circ f$.
}
\end{figure}

\subsection{Function composition}

\begin{definition}[composition]
\label{compDef}
Let $f \in S \rightarrow T$ and $g \in T \rightarrow U$.
The \emph{composition} $g \circ f$ of $f$ and $g$ is the function in
$S \rightarrow U$ that has as map $x \mapsto g(f(x))$.
\end{definition}

See Figure~\ref{fig:fourExamples}.
We sometimes write $g \circ f$
without explicitly saying that it is defined;
that is, that the target of $f$ is the source of $g$.

\begin{example}
For all $f \in S \rightarrow T$
we have $f \circ id_S = f = id_T \circ f$.
Let $S'$ be a subset of $S$ and let $i$
be the insertion function in $S' \rightarrow S$.
Then $f \downarrow S' = f \circ i$.
See Figure~\ref{fig:fourExamples}.
\end{example}

The order of $f$ and $g$ in $g\circ f$
is derived from the expression $g(f(x))$ in the definition of composition.
In many situations this order seems unnatural.
If the $f$ and $g$ were functional binary relations,
then their composition would be written as $f;g$.

Function composition is associative:
if, in addition to Definition~\ref{compDef},
we have $h \in U \rightarrow V$,
then $(h \circ g) \circ f$ and $h \circ (g \circ f)$
are the same function in $S \rightarrow V$.
See Figure~\ref{fig:fourExamples}.

A useful property of composition
is that if $g\circ f$ is injective, then $f$ is.
For suppose that $f$ is not injective.
Then there exist $x_1$ and $x_2$ in $S$
such that $x_1 \not= x_2$ and $f(x_1) = f(x_2)$.
Then we also have that $g(f(x_1)) = g(f(x_2))$,
so that $g\circ f$ would not be injective.

The dual counterpart of this property is that
if $g\circ f$ is surjective, then $g$ is.
For suppose that $g$ is not surjective.
Then there is a $z$ in $U$
such that there is no $y$ in $T$ with $g(y) = z$.
Then there is also no $x$ in $S$ with $g(f(x)) = z$.

\begin{figure}[htbp]
\begin{center}
\epsfig{file=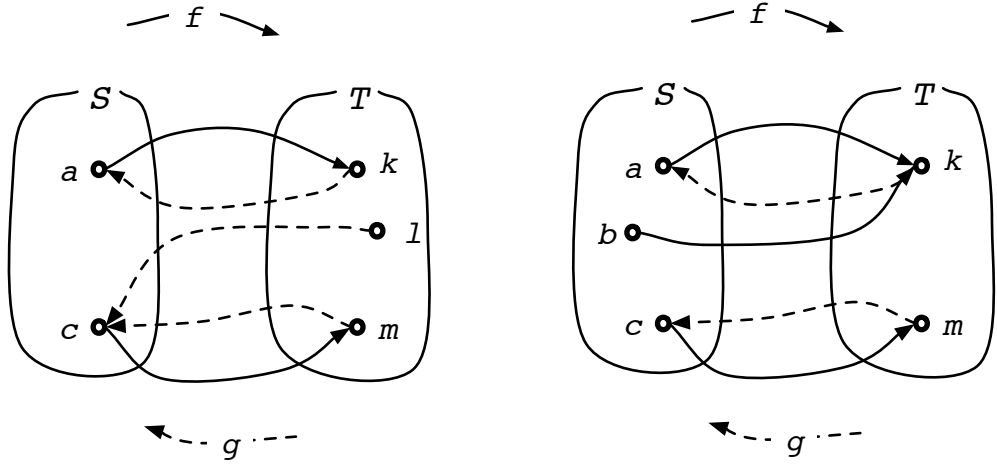}
\end{center}
\caption{\label{fig:leftRightInv}
On the left:
$g \circ f = \id_S$. For this, $f$ needs to be injective,
but need not be surjective. As it is not, $g$ is not uniquely
determined: $g(l)$ could be $a$ and we would still have $g \circ f = \id_S$.
On the right:
$f \circ g = \id_T$. For this, $g$ needs to be injective,
but need not be surjective. As it is not, $f$ is not uniquely
determined: $f(b)$ could be $m$ and we would still have $f \circ g = \id_T$.
}
\end{figure}

\paragraph{Left and right inverse}
Let $S$ and $T$ be nonempty sets, let
$f \in S \rightarrow T$,
and let
$g \in T \rightarrow S$.
If $\id_S = g \circ f$, then
we say that $f$ is a \emph{right inverse} of $g$
and that
$g$ is a \emph{left inverse} of $f$.
In this case,
$f$ is injective and $g$ is surjective.

To see why,
note that $\id_S$ is injective and surjective.
Hence, by the property of composition just introduced,
$\id_S = g \circ f$ implies
that $f$ is injective and that $g$ is surjective.

If, in that case, $f$ is not surjective,
then there exists a $y$ in $T$ such that there does not exist
an $x$ in $S$ such that $f(x) = y$.
For such a $y$, the value of $g$ does not affect $g \circ f$.
Hence, $f$ and $\id_S = g \circ f$ do not
in general uniquely determine $g$.

But $\id_S = g \circ f$ and $f$ surjective do determine
$g$. In that case $f$ is a bijection, so that the inverse $f^{-1}$
exists. The unique $g = f^{-1}$ is a bijection as well, so
that $g$ is also injective.

Conversely, injectivity of $f \in S \rightarrow T$
implies the existence of a left inverse $g \in T \rightarrow S$ such that
$g \circ f = id_S$, which is then surjective.

\subsection{Set extensions of functions}
Whenever we have a function type $S \rightarrow T$,
it is natural to consider
the type $\Pwst(S) \rightarrow \Pwst(T)$.
Let $f$ be in $S \rightarrow T$.
We call a \emph{set extension} of $f$
any function $g \in \Pwst(S) \rightarrow \Pwst(T)$ such that
for all subsets $S'$ of $S$ we have that
$
\{f(s) \mid s \in S' \} \subset g(S').
$

This definition suggests a partial order among the set
extensions of a given $f \in S \rightarrow T$.
We define $g_0 \preceq g_1$ for set extensions
$g_0$ and $g_1$ of $f$ iff 
$g_0(S') \subset g_1(S')$ for all subsets $S'$ of $S$.
The relation $\preceq$ is a partial order.
In this partial order there is a least element.
Its map is $S' \mapsto \{f(x) \mid x \in S'\}$
for all subsets $S'$ of $S$.
There is a greatest element in the partial order.
Its map is $S' \mapsto T$
for all subsets $S'$ of $S$.

The least set extension of $f \in S \rightarrow T$,
the one that has as map
$S' \mapsto \{f(s) \mid s \in S' \}$,
is called the \emph{canonical set extension} of $f$. 
Conversely,
the function $h$ in $\Pwst(T) \rightarrow \Pwst(S)$
such that for all subsets $T'$ of $T$
$$
h(T') = \{x \in S \mid f(x) \in T' \}
$$
is the \emph{inverse set extension} of $f$.
Note that the inverse set extension is defined for any function,
whether it is a bijection or not.

Usually,
the canonical set extension of $f$
is just written as $f$,
and the inverse set extension as $f^{-1}$,
so that we write
$f(S') = \{f(x) \mid x \in S'\}$
and
$f^{-1}(T') = \{x \in S \mid f(x) \in T'\}$.
Conventional notation relies
on context to disambiguate $f^{-1}$,
because it may mean the inverse of a bijection $f$
or it may mean the inverse set extension
of an $f$ that is not necessarily a bijection.

An important property
that distinguishes some set extensions
is their \emph{monotonicity}\/:
$S_1 \subset S_2$ implies $f(S_1) \subset f(S_2)$.
Monotonicity gives us
$f(S_1 \cap S_2) \subset f(S_1)$
and
$f(S_1 \cap S_2) \subset f(S_2)$,
so that
$f(S_1)$ and
$f(S_2)$
are both supersets of
$f(S_1 \cap S_2)$.
As $f(S_1) \cap f(S_2)$ is the least common superset
of $f(S_1)$ and $f(S_2)$, we have
$$f(S_1 \cap S_2)       \subset   f(S_1) \cap f(S_2).$$
But the reverse inclusion does not in general hold.
For example, suppose that $f(x_1) = f(x_2) = y$ and $x_1 \not= x_2$
and let us take
$S_1 = \{x_1\}$
and
$S_2 = \{x_2\}$.
Then $y \in f(\{x_1\}) \cap f(\{x_2\})$,
whereas $y$ is not contained in $f(\{x_1\}\cap \{x_2\}) = f(\emptyset) = \emptyset$.
Injectivity of $f$ is a necessary and sufficient condition for
equality.

On the other hand, we do have
\begin{equation}\label{eq:cupInc}
f(S_1) \cup f(S_2)       =      f(S_1 \cup S_2)
\end{equation}
That the left hand side
is contained in the right hand side
follows from monotonicity and union being the least common superset.
For the reverse inclusion
suppose that $y \in f(S_1 \cup S_2)$.
Then there is an $x \in S_1 \cup S_2$
such that $f(x) \in f(S_1 \cup S_2)$.
If the $x$ is in $S_1$,
then $f(x)$ is in $f(S_1)$;
hence in $f(S_1) \cup f(S_2)$.
Otherwise, the $x$ is in $S_2$,
hence $f(x)$ is in $f(S_2)$;
hence in $f(S_1) \cup f(S_2)$.
Either way, $y \in f(S_1) \cup f(S_2)$.
This shows that the right hand side of (\ref{eq:cupInc})
is included in the left hand side.

For any $f$ in $S \rightarrow T$ we have
\begin{eqnarray*}
S' & \subset &  f^{-1}(f(S'))          \\
T' & \supset &  f(f^{-1}(T'))
\end{eqnarray*}
with $S'$ and $T'$ arbitrary subsets of $S$ and $T$.
In the first case we have equality iff $f$ is injective;
in the second case we have equality iff $f$ is surjective.

\subsection{Enriched sets}

$S$ and $T$ are the same set
if and only if every element of $S$ belongs to $T$ and vice versa.
Thus sets lack some properties, such as order,
that are sometimes desirable in a collection.
Moreover,
set membership is all or nothing, not a matter of degree.
Many authors consider the set-theoretic concept of a set inadequate.
This leads to various proposals for enrichment of the set concept.

The method of set theory is to use functions
to specify additional properties,
not to enrich the concept of set itself.

\paragraph{``Ordered sets''}
Suppose we wish to specify an order $\preceq$
among the elements of a finite set $S$.
We can do this by defining a bijection $f \in \iot{n} \rightarrow S$,
where $n \in \Nat$.
For every
$x_1 \in S$ and
$x_2 \in S$
there exist
$i_1 \in \iot{n}$ and
$i_2 \in \iot{n}$
such that
$f(i_1) = x_1$ and 
$f(i_2) = x_1$.
Then we can define
$x_1 \preceq x_2$ according to whether $i_1 \leq i_2$.

In set theory,
$S$ in combination with such an $f$
takes the place of an ``ordered set''. 

\paragraph{``Multisets''}
Sometimes one considers a collection where it is not enough
to say whether or not an element belongs,
but one wants to say how many times it belongs.
Such a collection is then called a \emph{multiset} or \emph{bag}.

In set theory such a requirement poses no difficulty.
It is a special case of the common situation that one wants
to associate an \emph{attribute} with each element of a set.
Suppose we have a set $S$ of objects and a set $T$ of attributes.
A function $f \in S \rightarrow T$ then specifies for
each $x \in S$ which of the attributes $f(x) \in T$ it has.
If $T = \Nat$, then the attribute $f(x)$ can be regarded as
the multiplicity of $x$.

In set theory,
$S$ in combination with such an $f$
takes the place of a ``multiset''. 

\paragraph{``Fuzzy sets''}
Sometimes one considers a collection where it is not enough
to say whether or not an element belongs,
but one wants to say how strongly it belongs.

In set theory, one would regard
the required strength of belonging as an attribute.
The interval $[0,1]$ of real numbers
can be selected as the set of attributes.
A set $S$ in combination with $f \in S \rightarrow [0,1]$
can then be regarded as a ``fuzzy set''.

\section{Tuples}
Often a function in $S\rightarrow T$
models a \emph{process} of which the input
and output consist of elements of $S$ and $T$, respectively.
A function $f \in S\rightarrow T$
may be used in a different way:
as a method for \emph{indexing} elements of $T$
using the elements of $S$ as index.

When $f$ is used in this way, then it is called
a \emph{family} or a \emph{tuple} with $S$ as \emph{index set},
of which the components are restricted to belong to $T$.

In such a situation one sometimes writes
$f_s$ rather than $f(s)$ for the unique element
of $T$ indexed by $s \in S$.
Arrays in programming languages are examples of such families.
Then $S = \iot{n}$ and one would write $f[i]$
rather than $f(i)$ where $i \in \iot{n}$.

If the index set is numerical,
as in $S = \iot{n}$, for some $n \in \Nat$, or $S = \Nat$,
then a tuple in $S \rightarrow T$ is called a \emph{sequence}.
If $S = \Nat$, then the sequence is said to be \emph{infinite}.
If $S = \iot{n}$, then we call $n$
the \emph{length} of the sequence.

If we think of $T$ as an alphabet,
then the functions in $\iot{n} \rightarrow T$
are the \emph{strings over} $T$ \emph{of length} $n$.
Suppose we have strings
$\alpha \in \iot{m} \rightarrow T$
and
$\beta \in \iot{n} \rightarrow T$.
Then
$\gamma \in (m+n) \rightarrow T$
is the \emph{concatenation} of $\alpha$ and $\beta$
if the map of $\gamma$ is given by
$i \mapsto \alpha(i)$ if $0 \leq i < m$
and
$i \mapsto \beta(i-m)$ if $m \leq i < m+n$.

\subsection{Notation}

Parentheses in mathematics are overloaded with meanings.
There is the prime purpose of indicating tree structure in expressions.
An unrelated meaning is to indicate function application,
as in $f(x)$, although these two meanings seem to combine in $f(x+y)$.
And then we see that a sequence
$x$ in $\iot{n} \rightarrow T$
is often written as $(x_0, \ldots, x_{n-1})$.
Let us at least get rid of this last variant by writing
$x \in \iot{n} \rightarrow T$ as
$\langle x_0, \ldots, x_{n-1}\rangle$\footnote{
A potential pitfall of this notation for tuples is that
$\pair{a}{b}$ can now mean either a pair
or a tuple $p$ with $\iot{2}$ as index set
such that $p_0 = a$ and $p_1 = b$.
}.

Tuples need not be indexed by integers.
For example, points in the Euclidean plane are reals indexed by
the $X$ and $Y$ coordinates.
Describing points this way corresponds  to regarding a point $p$
as a tuple in $\{X,Y\} \rightarrow \Rea$.
For example, a point $p$ could be such that $p_X = 2$ and $p_Y = 3$.

In a context where the points are thought of as being characterized by
the coordinates $X$ and $Y$ one often sees
$p=\pair{2}{3}$.
Apparently, a quick switch has been made
from $\{X,Y\}$
to $\{0,1\}$ as index set.
We need a notation that unambiguously describes tuples
with small nonnumerical index sets.

Consider tuples in $S \rightarrow T$ representing addresses.
Here the index set $S$ could be
$$
\{
\mbox{number},
\mbox{street},
\mbox{city},
\mbox{state},
\mbox{zip}
\}.
$$
Such tuples can be written as lines in a table
with columns labeled by the elements of the index set.
The ordering of the columns is immaterial.

In this way addresses
$a$,
$b$,
$c$,
$d$, and
$e$
would be described as shown in Figure~\ref{fig:addresses}.

\begin{figure}[htbp]
\begin{center}
\begin{tabular}{l||r|l|l|l|l}
     & number &  street & city & state & zip      \\
\hline
$a$  & 19200  &  120th Ave                 & Bothell     & WA & 98011    \\
$b,c$  & 2200   &  Mission College Boulevard & Santa Clara & CA & 95052   \\
       & 2201   &  C Street NW               & Washington  & DC & 20520  \\
$d$  & 36     &  Cooper Square             & New York    & NY & 10003   \\
$e$  & 2201   &  C Street NW               & Washington  & DC & 20520
\end{tabular}
\end{center}
\caption{\label{fig:addresses}
A sample of named and unnamed tuples.
}
\end{figure}

In Figure~\ref{fig:addresses},
the area to the left of the double vertical line
is not a column: it only serves to write the names, if any, of the tuples.
The third tuple has no name; it is still a tuple.
There is a tuple that has two names: $b$ and $c$.
Another tuple occurs twice in the table, once with its name shown;
the other time without the name.

Without a table we would have to specify laboriously
$a_{\mbox{number}} = 19200$,
$a_{\mbox{street}} = \mbox{120th Ave}, \ldots$,
$b_{\mbox{number}} = 2200$,
$b_{\mbox{street}} = \mbox{Mission College Boulevard}$,
$\ldots$, and so on.
Such notation is unavoidable if tables are not available.
For example in XML\footnote{
The Extensible Markup Language, a standard promulgated by
the World Wide Web Consortium.
}
the tuple $a$ would be an ``element''
endowed with ``attributes'':
\begin{verbatim}
<a number = "19200"     street = "120th Ave"    city = "Bothell"
   state = "WA"         zip = "98011"
>
\end{verbatim}

If no tuple has a name, then one might as well omit the double vertical line,
and describe the four tuples by means of the table:
\begin{center}
\begin{tabular}{r|l|l|l|l}
      number &  street & city & state & zip      \\
\hline
      19200  &  120th Ave                 & Bothell     & WA & 98011    \\
      2200   &  Mission College Boulevard & Santa Clara & CA & 95052   \\
      36     &  Cooper Square             & New York    & NY & 10003   \\
      2201   &  C Street NW               & Washington  & DC & 20520  \\
\end{tabular}
\end{center}

In the same way, the point $p$ in $\{X,Y\} \rightarrow \Rea$
such that $p_X = 2$ and $p_Y = 3$
should be described by
\begin{center}
\begin{tabular}{l||l|l}
    & X  &  Y  \\
\hline
$p$ & 2  &  3 
\end{tabular},
\end{center}
instead of $p = \pair{2}{3}$,
which would incorrectly imply $\iot{2}$ as index set.
We could, of course,
code the $X$-coordinate as $0$ and the $Y$-coordinate as $1$,
so that it would be correct
to write $p$ as $\pair{2}{3}$.
The point is that we don't \emph{have} to code
familiar symbols such as $X$ and $Y$ for coordinates
as something else for the sake of set-theoretic modeling.

\subsection{Typed tuples}

Consider a tuple with
$
\{
\mbox{number},
\mbox{street}, 
\mbox{city},
\mbox{state},
\mbox{zip}   
\}
$
as index set, representing an address.
We would like to ensure that the elements indexed by
\emph{number} and \emph{zip} belong to \Nat\ and that the
others belong to the set \AlpStar\ of alphabetic character strings.
This is an example of \emph{typing} a tuple.
This involves in general associating a set with each index.
That is, to type a tuple with index $I$, we use a tuple
of sets with index $I$.

\begin{example}
Addresses could be typed by the tuple of sets

\begin{center}
\begin{tabular}{c|c|c|c|c}
number &  street & city & state & zip      \\
\hline
$\Nat$  &  $\AlpStar$ & $\AlpStar$ & $\AlpStar$ & $\Nat$
\end{tabular}
\end{center}

\end{example}

\begin{definition}[typing of tuples]
Let $I$ be a set of indexes, $T$ a set of disjoint sets,
$\tau$ a tuple in $I \rightarrow T$, and 
$t$ a tuple in $I \rightarrow \cup T$.
We say that $t$ is \emph{typed by} $\tau$ iff
we have $t(i) \in \tau(i)$ for all $i \in I$.
\end{definition}

\begin{figure}[htbp]
\begin{center}
\epsfig{file=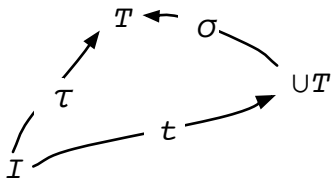}
\end{center}
\caption{\label{fig:typedTuple} 
The tuple $t \in  I \rightarrow \cup T$ is typed by
$\tau \in  I \rightarrow T$ iff $\tau = \sigma \circ t$.
}
\end{figure}

Because the sets in $T$ are disjoint,
there is a function $\sigma \in \cup T \rightarrow T$ that maps
each element $x$ of $\cup T$ to the set that is an element of $T$
to which $x$ belongs.
The condition of $t$ being typed by $\tau$ can be expressed by
means of function composition; see Figure~\ref{fig:typedTuple}.

\subsection{Cartesian products}

\begin{definition}[Cartesian products]
Let $I$ be a set of indexes, $T$ a set of disjoint sets, and
$\tau$ a tuple in $I \rightarrow T$.
The \emph{Cartesian product on} $\tau$,
denoted $\cart(\tau)$, is the set
of all tuples that have $\tau$ as type.
\end{definition}

Instead of `` $t$ is typed by $\tau$'',
we can now say `` $t \in \cart(\tau)$''.

\begin{example}\label{addressCP}
Let $\tau$ be in
$
\{
\mbox{number},
\mbox{street},
\mbox{city},
\mbox{state},
\mbox{zip}
\}
\rightarrow \{\Nat,\AlpStar\},
$
More specifically, let $\tau$ be equal to the tuple
\begin{center}
\begin{tabular}{c|c|c|c|c}
number &  street & city & state & zip      \\
\hline
\Nat & $\Alp^\ast$ & $\Alp^\ast$ & $\Alp^\ast$ & \Nat
\end{tabular}.
\end{center}

$\cart(\tau)$ is the set of all tuples
of type $\tau$. In this example it means that any natural number, however
small or large, can occur as the element indexed by \emph{number}
or \emph{zip}, and that any sequence of characters, in any combination,
of any length, can occur as element indexed by
\emph{street},
\emph{city}, or
\emph{state}.
The set of actually existing addresses
is a relatively small subset of $\cart(\tau)$.
\end{example}

\begin{example}
Let the index set $\{X,Y\}$
contain the labels that distinguish
the $X$ and $Y$ coordinates of the points in the Euclidean plane.
Let $\tau$ be the tuple of sets in $\{X,Y\} \rightarrow \{\Rea\}$.
Now $\cart(\tau)$ is the Euclidean plane.
\end{example}

Often the index set of $\tau$
is $\iot{n}$ for some positive natural number $n$.
For this case a special notation exists for $\cart(\tau)$.
Let $\tau$ be in $\iot{n} \rightarrow T$,
where $T$ is a set of sets.
The special nature of the index set allows us to
write $\tau$ as 
$\langle\tau_0,\ldots,\tau_{n-1}\rangle$.
In the same spirit,
$\cart(\tau)$ may be written as
$\tau_0 \times \cdots \times \tau_{n-1}$.
When
$T$ consists of a single set, say, $T = \{S\}$,
then
$\tau \in \iot{n} \rightarrow \{S\}$,
and
$\tau_0 \times \cdots \times \tau_{n-1}$
may be written as
$\underbrace{S \times \cdots \times S}_{n\; times }$,
abbreviated as $S^n$.

\begin{example}
$S \times T = \cart(\tau)$
where $\tau \in \iot{2} \rightarrow \{S,T\}$ with
$\tau(0) = S$ and $\tau(1) = T$.
I trust that the context prevents confusion with
the binary Cartesian product $S \times T$ defined as
$\{\pair{x}{y} \mid x \in S \wedge y \in T\}$
(see Definition~\ref{binCart}).
\end{example}

\begin{example}
A pack of playing cards can be modeled as a Cartesian product.
Let $\mbox{\textsc{suit}} = \{
\clubsuit,     
\diamondsuit,
\heartsuit,
\spadesuit
\}$ and
let $\mbox{\textsc{value}} = \{2,3,4,5,6,7,8,9,10,J,Q,K,A\}$.
Let $\tau = \pair{\textsc{suit}}{\textsc{value}}$.
Then the 52 elements of $\cart(\tau)$,
which include e.g.  
$\pair{\heartsuit}{Q}$, and
$\pair{\spadesuit}{5}$,
can be interpreted as playing cards.
As the index set of $\tau$ is $\iot{2}$,
we can write $\cart(\tau)$ as
$\mbox{\textsc{suit}}\times \mbox{\textsc{value}}$.
\end{example}

\begin{example}
Let the index set $I = \iot{n}$ for some natural number $n$
and let $T = \{\Rea\}$.
Then $I \rightarrow T$
consists of one typing tuple, say $\tau$.
$\cart(\tau)$
consists of the points in Euclidean $n$-space
and we write $\Rea^n$ instead of $\cart(\tau)$.
\end{example}

\subsection{Functions and Cartesian products}

We have considered functions $f$ in $S \rightarrow T$.
These have one argument $x$ in $S$ that gets mapped into $f(x) \in T$.
$S$ or $T$, or both, may be a Cartesian product.

Suppose
$S = \cart(\tau)$, with $\tau \in I \rightarrow U$.
The argument of $f$ is a tuple that is not necessarily indexed by numbers.
However, if $I = \iot{n}$,
then we have $f \in U_0 \times \cdots \times U_{n-1} \rightarrow T$
and $f$-values look like
$f(\langle u_0, \ldots, u_{n-1}\rangle)$,
usually written as
$f(u_0, \ldots, u_{n-1})$,
where
$u_0 \in U_0, \ldots, u_{n-1} \in U_{n-1}$.
This is an alternative to Currying for modeling multi-argument
functions.

It can also happen that the target set $T$
of a function $f$ in $S \rightarrow T$
is a Cartesian product.
In such a case it may be more natural to think of $f$ as a tuple $f'$
of functions.
Suppose that $f$ is in $S \rightarrow \cart(\tau)$
with $\tau \in I \rightarrow U$, where $U$ is a set of sets.
Thus, for each $x \in S$, we have that $f(x)$ is a tuple with
index set $I$.
For each $i \in I$, $(f(x))_i$ is an element of this tuple.

One can also think of such an $f$
as a tuple $f'$ of functions with index set $I$.
For each $i \in I$,
we define the element $f'_i$ of the tuple $f'$
as the function in $S \rightarrow \tau_i$
with map $x \mapsto (f(x))_i$.

For every function $f$ that has a Cartesian product as target,
there is a tuple $f'$ of functions uniquely determined as shown above.
Conversely, for every tuple of functions that have the same source,
the above shows the uniquely determined function with this source
that has tuples as values.

\subsection{Projections}
Consider a sequence $s = \langle a,b,c,d,e \rangle$
in $\iot{5} \rightarrow \Alp$,
where \Alp\ is a set of alphabetical characters.
Which, if any, of the following is a ``subsequence'' of $s$:
$\triple{b}{c}{d}$,
$\triple{a}{c}{e}$,
$\triple{e}{c}{a}$, or
$\triple{a}{a}{c}$?
Authors' opinions differ.

We saw that the definition of function leads in a natural way
to the notion of restriction.
Sequences, being tuples, being functions, therefore also have restrictions.
In set theory, subtuples are modeled as restrictions:
\begin{definition}[subtuple]
If $S' \subset S$, then $f \downarrow S'$
is the \emph{subtuple} on $S'$ of the tuple
$f$ in $S \rightarrow T$.
\end{definition}

\begin{figure}[htbp]
\begin{center}
\epsfig{file=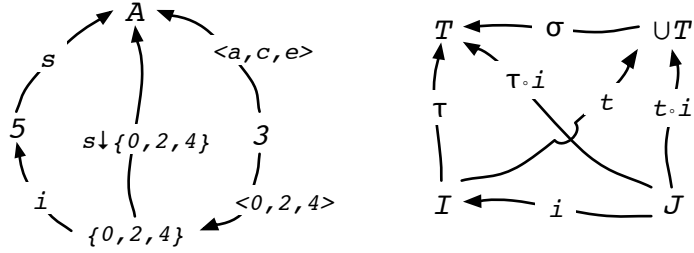}
\end{center}
\caption{\label{fig:tuples}
Left: See Example~\ref{ex:subTuple}.
Right: See Definition~\ref{def:subType}.
The fact that $t$ is typed by $\tau$
is expressed by the fact that $\tau = \sigma \circ t$.
It follows that $\tau \circ i = \sigma \circ t \circ i$.
}
\end{figure}

\begin{example}
\label{ex:subTuple}
Let $s = \langle a,b,c,d,e \rangle$.
We could restrict
the index set of $s$, which is $\iot{5}$,
to $S' = \{0,2,4\}$
and get as a subtuple of $s$
the restriction $s \downarrow S'$, which is
\begin{tabular}{l|l|l}
0  &  2 & 4  \\
\hline
a  &  c & e
\end{tabular},
and which is not
$\triple{a}{c}{e}$.
In fact, we have that
$\triple{a}{c}{e} = (s \downarrow S')\circ \triple{0}{2}{4}$,
if the target of $\triple{0}{2}{4}$ is $\{0,2,4\}$.
See Figure~\ref{fig:tuples}.
\end{example}

Though tuples are functions, we saw that they come with their
own terminology.
We saw that it is ``index set'' instead of ``source'' and ``subtuple''
instead of restriction.
In fact, there is additional specialized terminology for
tuples.

\begin{definition}[projection]
\label{def:tupleProj}
Let $t$ be a tuple
with index set $I$ and let $J$ be a subset of $I$.
The \emph{projection} $\pi_J(t)$ of $t$ on $J$
is the subtuple $t \downarrow J$ of $t$.
\end{definition}

Thus we have
$\pi_{\{0,2,4\}}(\langle a,b,c,d,e \rangle) =
\begin{tabular}{l|l|l}
0  &  2 & 4  \\
\hline
a  &  c & e
\end{tabular}
$.

When the projection is on a singleton set of indexes,
the result is a singleton tuple.
Rather than writing
$\pi_{\{3\}}(\langle a,b,c,d,e \rangle) = 
\begin{tabular}{l}
3 \\
\hline
d
\end{tabular}
$,
we may write the singleton set of indexes as the index by itself:
$\pi_3(\langle a,b,c,d,e \rangle) =
\begin{tabular}{l}
3 \\
\hline
d
\end{tabular}
$.

If a tuple $t$ has type $\tau$,
then the subtuples of $t$ are typed in an obvious way.
This observation suggests the definition below.
\begin{definition}[subtype]
\label{def:subType}
Let $T$ be a set of disjoint sets,
$I$ an index set, $J$ a subset of $I$,
and $\tau$ a type in $I \rightarrow T$.
We define $\tau \downarrow J$
to be the \emph{subtype} of $\tau$ determined by $J$. 
\end{definition}
Let now $t \in I \rightarrow \cup T$
be a tuple typed by $\tau$.
It is easy to see that the subtuple $t \downarrow J$ of $t$
is typed by the subtype $\tau \downarrow J$ of $\tau$.
Observe that, if $i \in J \rightarrow I$ is the insertion function,
then $t \downarrow J = t \circ i$.
In Figure~\ref{fig:tuples} we see that
$\tau \downarrow J = \sigma \circ (t \downarrow J)$.
This shows that if we want to say
that $t \downarrow J$ is typed by $\tau \downarrow J$,
then we can say it with arrows.

A projection that is defined for an individual tuple
is also defined, as canonical set extension,
on a \emph{set} of tuples that have the same type.
Consider for example a Cartesian product.
It is a set of tuples of the same type.
Therefore
$\pi_J(\cart(\tau))$ is defined by canonical set extension
to be equal to
$\{\pi_J(t) \mid t \in \cart(\tau)\}$.
These are all the tuples of type 
$\pi_J(\tau)$, so by definition equal to 
$\cart(\pi_J(\tau))$.
Thus we see that a projection of a Cartesian product
is itself a Cartesian product.

\section{Relations}

In Section~\ref{binRelSec} we defined binary relations.
In this section we define a kind of relation that can hold
between any number of arguments.
We call these ``relation''. Whenever we refer to a \emph{binary} relation,
we should always qualify it as such.
The generality of relations implies that some of them hold
between two arguments. But these are still relations and not binary relations.
The situation is similar to the trees in Knuth's Art of Programming \cite{knth68}.
He defines binary trees and trees in such a way that binary trees
are not a special case of trees.

Relations are specified by means of sets
consisting of the tuples containing the objects that are in the relation.
It is not enough for the sizes of these tuples to be the same.
Just as the source and the target
are parts of the specification of a function,
a relation is most useful as a concept
when its tuples have the same type.
Therefore this type is part of the specification of a relation.

\begin{definition}[relation]\label{relDef}
A \emph{relation} is a pair $\pair{\tau}{E}$
where $\tau$, the \emph{signature}, or the \emph{type}, of the relation,
is a tuple of type $I \rightarrow T$, where $T$ is a set of sets,
and $E$, the \emph{extent} of the relation,
is a subset of $\cart(\tau)$.
\end{definition}

\begin{example}
$\pair{\tau}{E}$ where $\tau$ has the empty set as index set.
$\cart(\tau)$ has one element, the empty tuple.
There are two possibilities for $E$:
the empty set and the singleton set containing the empty tuple.
\end{example}

The following example emphasizes the distinction
between a relation and a binary relation.
\begin{example}
A relation $\pair{\tau}{E}$ with $\tau \in \iot{2} \rightarrow T$
can be represented without loss of information by the binary relation
$\triple{\tau(0)}{\tau(1)}{\{\pair{t_0}{t_1}\mid t \in E\}}$.
Conversely, let $\triple{S}{T}{E}$ be a binary relation and let
$\tau \in \iot{2} \rightarrow \{S,T\}$ be such that
$\tau_0 = S$
and
$\tau_1 = T$.
This binary relation is represented without loss of information as
the relation $\pair{\tau}{F}$
where $F = \{t \in \cart(\tau) \mid \pair{t_0}{t_1} \in E  \}.$
\end{example}

The following example shows that set theory
is of potential interest to databases.
\begin{example}
Let $\tau$ be as in Example~\ref{addressCP}.
The actually existing addresses
are but a relatively small subset $E$ of $\cart(\tau)$.
The relation $\pair{\tau}{E}$ is a relational format
for the information represented by this set of addresses.
\end{example}

\begin{example}
The Euclidean plane is a set of tuples that have index set
$\{X,Y\}$ and where both elements are reals.
That is, the Euclidean plane is $\cart(\tau)$,
where $\tau$.

A figure in the Euclidean plane is
a subset of the Euclidean plane.
So every figure in the Euclidean plane is a relation of type
$\{X,Y\} \rightarrow \{\Rea\}$.
For example, the relation
$$
\pair{\tau}{\{p \in \cart(\tau) \mid p_X^2 + p_Y^2 = 1\}}
$$
is the unit circle with the origin as centre. 
\end{example}

\subsection{Set-like operations}
Among relations of the same type
certain relational operations are defined
that mirror set operations:

\begin{eqnarray*}
\pair{\tau}{E_1} \cup \pair{\tau}{E_2} & \stackrel{def}{=} &
              \pair{\tau}{E_1\cup E_2}         \\
\pair{\tau}{E_1} \cap \pair{\tau}{E_2} & \stackrel{def}{=} &
              \pair{\tau}{E_1\cap E_2}         \\
\pair{\tau}{E_1} \setminus \pair{\tau}{E_2} & \stackrel{def}{=} &
              \pair{\tau}{E_1\setminus E_2}
\end{eqnarray*}

\subsection{Relations with named tuples}

The extent of a relation
$\pair{\tau}{E}$ is the set of tuples $E$.
These tuples are not named or ordered.
Naming of the tuples can be achieved by a set $S$
of names and a function
$f$ in $S \rightarrow E$.
If we want to name \emph{all} tuples,
$f$ needs to be surjective.

It is also be possible to name the tuples of a relation
by means of a subset $I'$ of $I$ (where $I$ is the index set of $\tau$),
if $t\downarrow I'$ uniquely identifies $t$.
That is, if
$t_0\downarrow I' = t_1\downarrow I'$
implies, for all $t_0, t_1 \in E$, that $t_0 = t_1$.
Sometimes such an $I'$ is a singleton set $\{i\}$.
A common example is where $E$ consists of tuples $t$ describing
employees and $t_i$ is the social insurance number.

\subsection{Projections and cylinders}
Recall that projections are defined (Definition~\ref{def:tupleProj})
as restrictions of tuples.
Canonical set extensions of projections
are therefore defined on \emph{sets} of tuples.
As extents of relations are sets of tuples,
projections are also defined, as canonical set extensions,
on extents of relations.

\begin{definition}[projection]
\label{def:proj}
Let $\tau$ be in $I \rightarrow T$,
where $T$ is a set of disjoint sets
and $I$ is an index set.
The \emph{projection}
\emph{on} $J$ of the relation
$\pair{\tau}{E}$ is
written $\pi_J(\pair{\tau}{E})$
and is defined to be
the relation $\pair{\tau\downarrow J}{\{t \downarrow J \mid t \in E \}}$.
\end{definition}

\begin{example}
If we have a relation $r = \pair{\tau}{E}$
where $\tau$ is
\begin{center}
\begin{tabular}{c|c|c|c|c}
number &  street & city & state & zip      \\
\hline
$\Nat$  &  $\AlpStar$ & $\AlpStar$ & $\AlpStar$ & $\Nat$
\end{tabular}
\end{center}
and $E$ is the set of tuples that represent addresses of
subscribers, then
$\pi_{\{\mbox{city},\mbox{state}\}}(r)$
is a relational format for the cities where
there is at least one subscriber.
\end{example}

\begin{example}\label{aabb}
Let $r$ be the relation $\pair{\tau}{E}$
where $\tau = \iot{3} \rightarrow \{\{a,b\}\}$
and
$ E =
\{
\triple{a}{a}{a},
\triple{a}{a}{b},
\triple{b}{a}{b}
\}
$.

An example of a projection is
$$
\pi_{\iot{2}}(r) =
\pair
{\iot{2} \rightarrow \{\{a,b\}\}}
{\{\pair{a}{a},\pair{b}{a}\}}
$$
We cannot represent the extent of, e.g.,
$\pi_{\{0,2\}}(r)$ with angle brackets because the angle brackets presuppose
an index set in the form of $\iot{n}$.
Instead we say,
$$
\pi_{\{0,2\}}(r) =
\pair
{\{0,2\} \rightarrow \{\{a,b\}\}}
{
E
}
$$
where
$E = 
\begin{array}{l|l}    
0  &  2    \\
\hline
a  &  a    \\
a  &  b    \\
b  &  b
\end{array}
$.
\end{example}

Projections are more often used as canonical set extensions
on sets of tuples than on individual tuples.
As projections are typically not injective,
they typically do not have an inverse.
But every function does have an inverse set extension.
The inverse set extensions of projections are especially useful.

Let $I$ be an index set with a subset $J$.
If $S$ is a set of tuples with index set $J$,
then we can ask:
What tuples $t$ with index set $I$
are such that $\pi_J(t) \in S$?
This is the definition of inverse set extension:
it is defined for all functions, projections or not.
So
$$\pi^{-1}_J(S) = \{t \mid \pi_J(t) \in S\}. $$

\begin{example}
In Example~\ref{aabb},
$\pi^{-1}_{\{0,1\}}$
applied to the extent of
$\pi_{\{0,1\}}(r)$
and
$\pi^{-1}_{\{0,2\}}$
applied to the extent of
$\pi_{\{0,2\}}(r)$
are, respectively,
\begin{eqnarray*}
& &
\{
\triple{a}{a}{a},
\triple{a}{a}{b},
\triple{b}{a}{a},
\triple{b}{a}{b}
\}
\\
& &
\{
\triple{a}{a}{a},
\triple{a}{b}{a},
\triple{a}{a}{b},
\triple{a}{b}{b},
\triple{b}{a}{b},
\triple{b}{b}{b}
\}
\end{eqnarray*}
\end{example}

Given a relation $r$, one can also wonder about relations
that have $r$ as projection.
There is a largest such, which is the ``cylinder'' on $r$.
This gives the idea;
to get the signatures right, see the following definition.

\begin{definition}[cylinder]
For $i \in \{0,1\}$,
let $\tau_i$ be in $I_i \rightarrow T_i$,
such that $\tau_0$ and $\tau_1$ are summable;
$T_i$ is a set of sets.
Let $\pair{\tau_0}{E_0}$ be a relation.
The \emph{cylinder}
\emph{in} $I_0\cup I_1$
\emph{on} $\pair{\tau_0}{E_0}$
is written as
$\pi^{-1}_{I_0\cup I_1}(\pair{\tau_0}{E_0})$
and is defined to be relation
$$\pair{\tau_0+\tau_1}{\{t \in \cart(\tau_0+\tau_1) \mid t\downarrow I_0 \in E_0\}}.$$
\end{definition}
The notation $\pi^{-1}$ suggests some kind of inverse of projection.
The suggestion is inspired by facts such as
$\pi_{I_0}(\pi^{-1}_{I_0\cup I_1}(\pair{\tau_0}{E_0})) =
\pair{\tau_0}{E_0}$.

\begin{example}
Consider
$\cart(\tau)$, with $\tau$ in $\{X,Y,Z\} \rightarrow \{\Rea\}$,
and think of it as the three-dimensional Euclidean space.
Let the relation
$r$ be $\pair{\tau}{\{t \in \cart(\tau) \mid t_X^2+t_Y^2+t_Z^2 \leq 1\}}$,
which is the unit sphere with centre in the origin.
The projections 
\begin{eqnarray*}
\pi_{\{X,Y\}}(r)  & = &
\pair{\tau \downarrow \{X,Y\}}
       {\{t \in \pi_{\{X,Y\}}\cart(\tau) \mid t_X^2+t_Y^2 \leq 1\}}  \\
\pi_{\{Z,Y\}}(r)  & = &
\pair{\tau \downarrow \{Z,Y\}}
       {\{t \in \pi_{\{Z,Y\}}\cart(\tau) \mid t_Z^2+t_Y^2 \leq 1\}}  \\
\pi_{\{Z,X\}}(r)  & = &
\pair{\tau \downarrow \{Z,X\}}
       {\{t \in \pi_{\{Z,X\}}\cart(\tau) \mid t_Z^2+t_X^2 \leq 1\}}
\end{eqnarray*}
can be thought of as the two-dimensional projections of the sphere.

The one-dimensional projections
\begin{eqnarray*}  
\pi_X(r)  & = &
\pair{\tau \downarrow \{X\}}
       {\{t \in \pi_X\cart(\tau) \mid t_X^2 \leq 1\}}  \\
\pi_Y(r)  & = &
\pair{\tau \downarrow \{Y\}}
       {\{t \in \pi_Y\cart(\tau) \mid t_Y^2 \leq 1\}}  \\
\pi_Z(r)  & = &
\pair{\tau \downarrow \{Z\}}
       {\{t \in \pi_Z\cart(\tau) \mid t_Z^2 \leq 1\}}
\end{eqnarray*}
can be thought of as the one-dimensional projections of the sphere.

The cylinder
$
\pi_X^{-1}(\pi_X(r))
$
looks like an infinite slab bounded by two planes parallel
to the $Y,Z$-plane through the points
with $Y$ and $Z$ coordinates 0 and with $X$ coordinates $-1$ and $+1$.
This slab is just wide enough to contain the sphere.
$$
\pi_X^{-1}(\pi_X(r)) \cap
\pi_Y^{-1}(\pi_Y(r)) \cap
\pi_Z^{-1}(\pi_Z(r))
$$
is the intersection of three such slabs perpendicular to
each other, so it is the smallest cube with edges parallel to the coordinate
axes that contains the sphere.
It is a Cartesian product.

The cylinder
$\pi_{\{X,Y\}}^{-1}(\pi_{\{X,Y\}}(r))$
looks like, well, a \emph{cylinder}.
Its axis is parallel to the $Z$ axis.
The unit disk in the $X,Y$-plane
with its centre at the origin is a cross-section.
$$
\pi_{\{X,Y\}}^{-1}(\pi_{\{X,Y\}}(r)) \cap
\pi_{\{Z,X\}}^{-1}(\pi_{\{Z,X\}}(r)) \cap
\pi_{\{Z,Y\}}^{-1}(\pi_{\{Z,Y\}}(r))
$$
is the intersection of three such cylinders.
It is a body bounded by curved planes that
properly contains the sphere $r$ and is properly contained
in the box.
\end{example}

\begin{definition}[join]
\label{def:join}
For $i \in \{0,1\}$ let there be relations $\pair{\tau_i}{E_i}$,
with $\tau_i \in I_i \rightarrow T_i$ and $T_i$ a set of disjoint sets.
If
$\tau_0$
and
$\tau_1$
are summable, then the \emph{join}
of
$\pair{\tau_0}{E_0}$ and $\pair{\tau_1}{E_1}$
is written as
$\pair{\tau_0}{E_0} \Join \pair{\tau_1}{E_1}$
and defined to be
$
\pi^{-1}_{I_0 \cup I_1}(\pair{\tau_0}{E_0})
\cap
\pi^{-1}_{I_0 \cup I_1}(\pair{\tau_1}{E_1})
$.
\end{definition}

The intersection in this definition is defined because of the assumed
summability of $\tau_0$ and $\tau_1$.
The signature of $\pair{\tau_0}{E_0} \Join \pair{\tau_1}{E_1}$
is $\tau_0 + \tau_1$.

\section{Application to relational databases}
\label{relDatMod}

Codd \cite{codd70} proposed to represent
the information in ``large banks of formatted data''
as a collection of relations.
This proposal has been so successful
that databases are ubiquitous
and that most of these conform
to Codd's relational model for data.
The success of the relational model
is due to the fact that its mathematical nature
has made more manageable
the complexity that would have prevented
the earlier models to support
the enormous growth that databases have experienced.

I have selected relational databases
as example of an application in computer science
where elementary set theory is useful.
This is because the notion of relation
is far from clear in the early database literature.
As examples I have selected Codd's original paper \cite{codd70}
and Ullman's widely quoted textbook \cite{llmn88}.

\subsection{The relational model according to Codd}

In Codd \cite{codd70} we find the following definition,
the original one, in section 1.3 ``A Relational View of Data'':
\begin{quote}
The term \emph{relation} is used here in its accepted mathematical sense.
Given sets $S_1,\ldots,S_n$ (not necessarily distinct),
$R$ is a relation on these $n$ sets if it is a set of $n$-tuples
each of which has its first element from $S_1$, its second
element from $S_2$, and so on.
We shall refer to $S_j$ as the $j$th \emph{domain} of $R$. \\
\end{quote}

Note that $S_j$, not $j$, is the domain.
Thus $S_i$ and $S_j$ may be the same set,
even though $i \not = j$.
The above quote continues with:

\begin{quote}
For expository reasons, we shall frequently make use of an array
representation of relations, but it must be remembered that this 
particular representation is not an essential part of the relational
view being expounded. An array which represents an $n$-ary relation
$R$ has the following properties:
\begin{enumerate}
\item
Each row represents an $n$-tuple of $R$.
\item
The ordering of rows is immaterial.
\item
All rows are distinct.
\item
The ordering of columns is significant --- it corresponds to
the ordering $S_1,\ldots,S_n$ of the domains on which $R$ is defined
(see, however, remarks below on domain-ordered and domain-unordered
relations).
\item
The significance of each column is partially conveyed by labeling
it with the name of the corresponding domain.
\end{enumerate}
\end{quote}

Codd gives as example of such an array
the one shown in Figure~\ref{CoddArr1}.
He observes that this example
does not illustrate why the order of the columns matters.
For that he introduces the one in Figure~\ref{CoddArr2}.
\begin{figure}[!htb]
\begin{center}
\rule{10cm}{0.005in}
\begin{tabular}{ccccc}
supply & (supplier & part & project & quantity)  \\
       &  1        & 2    & 5       & 17         \\
       &$\ldots$   &$\ldots$ &$\ldots$ & $\ldots$ \\
\end{tabular}
\caption{
\label{CoddArr1}
Codd's first example: domains all different.
}
\rule{10cm}{0.005in}
\end{center}
\end{figure}

\begin{figure}[!htb]
\begin{center}
\rule{10cm}{0.005in}
\begin{tabular}{cccc}
component & (part & part & quantity)  \\
          &  1    & 5    & 9         \\
          &$\ldots$&$\ldots$ &$\ldots$ \\
\end{tabular}
\caption{
\label{CoddArr2}
Codd's second example: domains not all different.
}
\rule{10cm}{0.005in}
\end{center}
\end{figure}

He explains Figure~\ref{CoddArr2} as follows.
\begin{quote}
$\ldots$ two columns may have identical headings (indicating
identical domains) but possess distinct meanings with respect
to the relation.
\end{quote}

We can take it that in Figure~\ref{CoddArr2}
we have $n = 3$, $S_1 = S_2 = \mbox{ part}$,
and $S_3 = \mbox{ quantity}$.
As $S_1, \ldots, S_n$ need not all be different,
columns can only identified by $\{1,\ldots , n\}$.

Codd goes on to point out that in practice $n$
can be as large as thirty
and that users of such a relation
find it difficult to refer to each column
by the correct choice among the integers $1,\ldots,30$.
According to Codd, the solution is as follows.

\begin{quote}
Accordingly, we propose that users deal, not with relations,
which are domain-ordered, but with \emph{relationships},
which are domain-unordered.
\end{quote}

One problem is the term ``domain-ordered''.
The term suggests that the relation is ordered
by the domains $S_1,\ldots,S_n$.
But, as Codd warns us, these sets are necessarily distinct.
If there are fewer than $n$ domains,
then they cannot order the relation.

Another problem with this passage
is the introduction of ``relationships''
as distinct from ``relations''.
Codd notes in his description
of the relational view of data that
``the term relation is used in its accepted mathematical sense''.
He would have had a hard time to find
an accepted mathematical sense
for the distinction between ``relation'' and ``relationship''.

Codd started out with the bold idea
that the data that need to be managed in practice
can be organized as relations ``in their accepted mathematical sense''.
Within one page he was forced to retract
from this promising start to get bogged down
in the murky area
of ``domain-ordered relations'' versus ``domain-unordered relationships''.

The cause of the difficulty
is that Codd seemed to regard
it as somehow unmathematical
for the index set $I$ to be anything
that differs from $\{1,\ldots,n\}$.
In this paper,
the exposition leading up to Definition~\ref{relDef}
is intended to ensure that one is not even tempted
to entertain such a misconception. 

\subsection{The relational model according to Ullman}

The reader may have thought that the difficulties
in Codd \cite{codd70} would soon be straightened out
as this original proposal became mainstream computer science,
and, as such, the subject of widely quoted textbooks.
Let us look at one of these, the one by J.D. Ullman \cite{llmn88}.

\cite{llmn88} introduces ``The Set-Theoretic Notion of a Relation''
(page 43), a notion there also called ``the set-of-lists'' notion
of a relation.
	This is distinguished from ``An Alternative Formulation of Relations'',
one that is called ``relation in the set-of-mappings sense''.

The set-list-lists notion of a relation is
any subset of the Cartesian product $D_1 \times \cdots \times D_k$,
where $D_1, \ldots, D_k$ are domains.
While Codd was careful to say that the domains need not be distinct,
\cite{llmn88} says that there are $k$ domains.
But this is probably not intended.
\cite{llmn88} uses \emph{attributes} $A_1,\ldots,A_k$ to name the columns of a tabular
representation of a relation.
\cite{llmn88} does not say whether these are distinct,
but that is  probably intended.

In the set-of-lists type of relation,
columns are named by the indexes $1,\ldots,k$.
Apparently, the attributes $A_1,\ldots,A_k$
are redundant comments on the columns.

Starting on page 44, in the ``Alternative Formulation''
section, \cite{llmn88} makes the observation that the attributes can be used
to index the domains instead of the indexes $\{1,\ldots,k\}$ in the 
set-of-lists type of relation.
When the indexes are attributes, one has a ``relation in the set-of-mappings sense''.
This kind of relation is illustrated with examples rather than defined.

\cite{llmn88} observes that in the practice of database use,
relations are sets of mappings.
This make one wonder why the other kind, with its redundant numerical indexes,
was introduced.
The answer comes on page 53, in the section on relational algebra.

\begin{quote}
Recall that a relation is a set of $k$-tuples for some $k$,
called the arity of the relation.
In general, we give names (attributes) to the components of tuples,
although some of the operations mentioned below, such as union,
difference, product, and intersection, do not depend on the names of the components.
These operations do depend on there being a fixed, agreed-upon order for
the attributes; i.e. they are operations on the list  style of tuples rather than
the mapping (from the attribute names to values) style.
\end{quote}

As we saw earlier, these operations do not depend on the index set of the type
$\tau$ of a relation $\pair{\tau}{E}$ having ``a fixed, agreed-upon order''
for its elements.

Definition~\ref{relDef} obviates the need for Codd's distinction
between ``domain-ordered'' and ``domain-unordered''
relations and for Ullman's distinction between
``sets-of-lists type relations'' and
``relations in the sets-of-mappings sense''.
Codd's idea of a relational format for data was a promising one,
but the special case of relations as subsets of
$D_1 \times \cdots \times D_n$ is too special.
It is perhaps not too late to revisit Codd's idea with relations according
to Definition~\ref{relDef}.

\subsection{A reconstruction of the relational model
according to set theory and logic}

Let us now see what the relational model for data
would look like to someone who knows some set theory
and who was only told the general idea of \cite{codd70}
without the details as worked out by Codd.
We first look how data are stored, then how they are queried.
Finally, in this section, we introduce a
relational operation that facilitates querying.

\subsubsection{Relations are repository for data}
We have to start with what is implicit
in the very idea of a database, relational or not.
The information to be stored in a database
concerns various aspects of things like employees in an organization,
parts of an airplane, books in library, and so on. 
The general pattern seems to be
that a database describes a collection of \emph{objects}.

There is no limit to the information
one can collect about an object as it exists in reality.
Hence one performs an act of abstraction
by deciding on a set of \emph{attributes}
that apply to the object
and one determines what is the \emph{value}
of each attribute for this  particular object.
A consequence of this abstraction
is that it cannot distinguish between objects
for which the attributes have the same value.
The set of attributes has to be comprehensive enough
that this does not matter for the purpose of the database.
That is, within the microcosm of the database,
one assumes that Leibniz's principle of Identity of Indiscernibles holds.

The foregoing is summarized in the first of the following points.
The remaining points constitute a reconstruction
of the main ideas of a relational database as suggested by the first point.

\begin{enumerate}
\item
A database is a description of a world populated by \emph{objects}.
For each object, the database lists the
\emph{values} of the applicable \emph{attributes}.
\item
The database presupposes a set of attributes, and for each attribute,
a set of allowable values for this attribute.
Such sets of admissible values are called \emph{domains}.
Let $A$ be the set of attributes and let $T$ be a set of disjoint domains.
As each attribute has a uniquely determined domain,
this information is expressed by a function, say $\tau$,
that is of type $A \rightarrow T$.
\item
Not all attributes apply to every object.
Hence, two objects may be similar in the sense
that the same set of attributes applies to both.
Let us say that objects that are similar in this sense
belong to the same \emph{class}.
Hence a class is characterized by the subset of $A$
that contains the attributes of the objects in the class.
\item
The description of each object
is an association of a value with each attribute
that is applicable to the object.
That is, the object is represented by a tuple
that is typed by $\tau \downarrow I$,
where $I$ is the set of attributes
of the class to which the object belongs.
\item
The objects of a class are described by tuples of the same type.
As tuples of the same type are a relation,
\emph{the class is a relation} of that type.
As objects need not belong to the same class,
a database consists in general of multiple relations
$
\pair{\tau\downarrow I_0}{E_0},
\ldots,
\pair{\tau\downarrow I_{n-1}}{E_{n-1}}
$,
where
$I_0, \ldots, I_{n-1}$
are the attribute sets of the classes.
\item
The life cycle of a database includes a design phase
followed by a usage phase.
In the design phase $A$, $T$, and
$\tau \in A \rightarrow T$
is determined,
as well as the subsets
$I_0, \ldots, I_{n-1}$
of $A$.
This is the database \emph{scheme}.
In the usage phase the extents
$E_0, \ldots, E_{n-1}$
are added and modified.
With the extents added, we have a database \emph{instance}.
\end{enumerate}

Note that the set of attributes of one class might  be included
in another. This suggests a hierarchy of classes.
Note also that nothing is said about the nature of the values 
that the attributes might take.
Values might be restricted to simple values like numbers or strings.
Or they could be tuples of one of the relations.

It would seem that whether to include these possibilities
in a relational database would be a matter of trading off
flexibility in modeling against simplicity and efficiency
in implementation.
Not so: these possibilities amount to an \emph{object-relational
database}, the desirability of which is subject to heated controversy.

\subsubsection{Queries, relational algebras, query languages}

According the relational model,
relations are not only used as format
for the data stored in a database,
but also as format for queries;
that is, for selections of data
to be retrieved from the database.
Thus the results of queries are relations.
These relations depend on those that are stored in the database
and must therefore be the result of operations on them.
Let us see how we can use the operations introduced so far.

\begin{figure}
\begin{center}
\parbox{2in}{
$\suppliers =$\\
\begin{tabular}{l|l|l} 
\sid & \sname & \city     \\
\hline
321  & \lee   & \tulsa    \\   
322  & \poe   & \taos     \\
323  & \ray   & \tulsa    \\
\end{tabular}
}
$\;\;\;$
\parbox{2in}{
$ \parts = $\\
\begin{tabular}{l|l|l|r} 
\pid & \pname & \sid  & \pqty   \\
\hline
213  & \hose   & 322  & 13  \\   
214  & \tube   & 321  & 6   \\
215  & \shim   & 322  & 18  \\
\end{tabular}
}
\end{center}
\begin{center}
\parbox{2in}{
$ \projects = $ \\
\begin{tabular}{l|l|r} 
\rid & \pid & \rqty   \\
\hline
132  & 215  & 2  \\   
133  & 214  & 11   \\
134  & 213  & 18  \\
\end{tabular}
}
\end{center}
\caption{\label{fig:citiesParts}
The relations for Example~\ref{ex:citiesParts}.
}
\end{figure}

\begin{figure}
\begin{verbatim}
SELECT PNAME, CITY
FROM SUPPLIERS, PARTS, PROJECTS
WHERE PARTS.PID = PROJECTS.PID AND SUPPLIERS.SID = PARTS.SID
      AND RQTY <= PQTY
\end{verbatim}
\caption{\label{fig:SQL}
An SQL query to the relations in Figure~\ref{fig:citiesParts}. 
}
\end{figure}
\begin{example}[Shim in Taos] \label{ex:citiesParts}
Consider the relations
shown in the tables in Figure~\ref{fig:citiesParts}.
Each table consists of a line of headings
followed by the line entries of the tables.
The line entries represent the tuples of the relations.
Each table has three such lines.

The following abbreviations are used.
In the \suppliers\ table,
\sid\ for supplier ID,
\sname\ for supplier name, and
\city\ for supplier city.
In the \parts\ table,
\pid\ for part ID,
\pname\ for part name, and
\pqty\ for part quantity on hand.
In the \projects\ table,
\rid\ for project ID and
\rqty\ for part quantity required.

We want to know the part names and cities
in which there is a supplier
with a sufficient quantity on hand
for at least one of the projects.
\end{example}

The relation specified
by the SQL query in Figure~\ref{fig:SQL}
can be specified with relations
according to Definitions~\ref{relDef},
and with projection and join
according to Definitions \ref{def:proj}
and \ref{def:join}, respectively.

Let the tables in Figure~\ref{fig:citiesParts}
be the relations
$\suppliers = \pair{\tau_0}{E_0}$,
$\parts = \pair{\tau_1}{E_1}$, and
$\projects = \pair{\tau_2}{E_2}$.
In addition, there is a relation in the query for which there is
no table, namely the less-than-or-equal relation.
Mathematically, there is no reason to treat it differently from
the relations stored in tables.
Hence, we also include
it as $\LEQ = \pair{\tau_3}{E_3}$.

The index sets of
$\tau_0$,
$\tau_1$,
$\tau_2$
are the sets of the column headings of the tables
for \suppliers, \parts, and \projects, respectively:
for
$\tau_0$ the index set is $\{\sid, \sname, \city\}$,
for
$\tau_1$ it is $\{\pid, \pname, \sid, \pqty\}$,
for
$\tau_2$ it is $\{\rid, \pid, \rqty\}$.
For
$\tau_3$ it is $\{\rqty, \pqty\}$.

The extents
$E_0$,
$E_1$, and
$E_2$
are as described in Figure~\ref{fig:citiesParts}.
Moreover,
$E_3 = \{t \in \cart(\tau_3) \mid t_{\rqty} \leq t_{\pqty}\}$.

Consider now the relation
\begin{equation}\label{eq:query1}
\pi_{\{\pname,\city\}}= \suppliers \Join \parts \Join \projects \Join \LEQ.
\end{equation}

For the relations to be joinable, the signatures
$\tau_0$,
$\tau_1$,
$\tau_2$, and
$\tau_3$ have to be summable.
That is, for any elements common to their source sets,
they have to have the same value.
For example, the source sets of
$\tau_0$ and
$\tau_1$ have \sid\
in common.
They both map \sid\ to its domain,
which is the set of supplier IDs.
Therefore  
$\tau_0$ and
$\tau_1$ are summable;
hence $\cities \Join \parts$ is defined (Definition~\ref{def:join}).
Similarly with the other joins in the expression (\ref{eq:query1}),
which has as value the relation described by the SQL query in
Figure~\ref{fig:citiesParts}.

In the query in Example~\ref{ex:citiesParts} 
every relation occurs at most once.
In the following example we consider 
a query where this is not the case.

\begin{example}[Mary, Alan]
\label{ex:parentChild}
In Figure~\ref{fig:parentChild} we show a table
specifying a relation consisting of tuples of two elements
where one is a parent of the other. 
It is required to identify pairs of persons
who are in the grandparent relation.
\end{example}

\begin{figure}
\begin{center}
$\pc=$
\begin{tabular}{l|l}
\parent & \child \\
\hline
\mary   & \john    \\
\john   & \alan     \\
\mary   & \joan    \\
\end{tabular}
\end{center}
\caption{\label{fig:parentChild}
Relation for Example~\ref{ex:parentChild}.
}
\end{figure}

What distinguishes this query
from the one in Example~\ref{ex:citiesParts} is
that the relations do not occur in the join as given,
but are derived from the given relation.
On the basis of the derived relations
we create one in which the pairs
are in the grandparent relation.
In one of the SQL dialects this would be:
\begin{verbatim}
SELECT PC0.PARENT, PC1.CHILD
FROM   PC AS PC0, PC AS PC1
WHERE  PC0.CHILD = PC1.PARENT
\end{verbatim}
In this query,
the derived tables are obtained via the linguistic device
of renaming \pc\ to $\pc_0$ and $\pc_1$.

\subsubsection{An additional relational operation}

Do we have to introduce a special-purpose language
to handle a simple query such as this one?
In the following we show how an additional relational operation,
``filtering'', is sufficient to handle,
in combination with projection and join,
not only this query, but more generally,
to write relations expressions
that mimic the queries
of a powerful query language such as Datalog \cite{mrwrrn88}.

The filtering operation acts
on a relation and a tuple and results in a relation.
Let the relation be $\pair{\tau}{E}$
with $\tau \in I \rightarrow T$ and $T$ a set of
disjoint sets.
Let the tuple be $p \in I \rightarrow V$,
where $V$ is a set of objects that we think of as placeholders.
In similar situations such placeholders
are often called ``variables'',
which is fine, as long as we remember
that they are elements of a set,
and that there is nothing linguistic about them.

\begin{figure}[htbp] 
\begin{center}
\epsfig{file=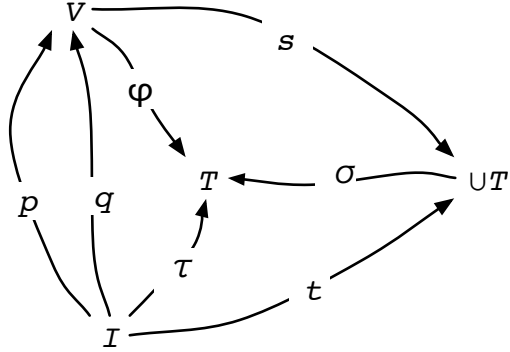} 
\end{center}
\caption{\label{fig:filter}
Sets and functions involved in filtering.
}
\end{figure}

For each such tuple $p$ we can ask:
is there a $t \in E$ such that $t = s \circ p$
for some $s \in V \rightarrow \cup T$?
In that case we include $s$ in the extent $E'$ of a relation
$\pair{\varphi}{E'}$.
See Figure~\ref{fig:filter}.

\begin{definition}[filtering]
\label{def:filter}
Let $\tau$ be in $I \rightarrow T$,
where $T$ is a set of disjoint sets
and $I$ is an index set.
Let $p \in I \rightarrow V$ be a tuple.
Let $\varphi \in V \rightarrow T$
be such that $\tau = \varphi \circ p$.
The \emph{filtering}
\emph{by} $p$ of the relation
$\pair{\tau}{E}$ is
written $\pair{\tau}{E}:p$
and is defined to be
the relation
$\pair{\varphi}
      {\{s \in p(I) \rightarrow \cup T
       \mid \exists t \in E.\; t = s \circ p
       \}}$.
\end{definition}

The condition 
$\tau = \varphi \circ p$ ensures that the variables
in $V$ are typed compatibly with $\tau$.

\begin{example}
\label{ex:square}
Let $I=3$, $T = \{\Rea\}$, $V = \{x,y,z\}$,
$\tau \in I \rightarrow T$,
and
$E = \{\triple{u}{v}{w} \in \Rea^3 \mid u \times v = w\}$.
Let $p = \triple{x}{x}{z}$.
What is $\pair{\tau}{E}:p$?
\end{example}

In this example we use the filtering operation
to obtain the squaring relation as a special case of the multiplication
relation.
As $T$ is a singleton set,
there is only one possibility for $\tau$,
and this is $\triple{\Rea}{\Rea}{\Rea}$.
The same holds for $\varphi$.
In this way, $\pair{\tau}{E}$ is the multiplication relation
over the reals.

Definition~\ref{def:filter} now implies
that the extent of $\pair{\tau}{E}:p$,
the result of applying filtering with $p$
to $\pair{\tau}{E}$,
is $\{s \in \{x,z\} \rightarrow \Rea \mid s_x^2 = s_z\}$.
This can be explained as follows.
We need to find all $s \in \{x,z\} \rightarrow \Rea$ such that there
exists a $t \in E$ with $t = s \circ p$.
Now, $t = s \circ p$ implies that
$t_0 = (s \circ p)_0 = s _{p_0} = s_x$.  
Similarly,
$t_1 = (s \circ p)_1 = s _{p_1} = s_x$.  

That is, $t = s \circ p$ has a solution $s$
iff $t_0 = t_1$.
Indeed, $p = \triple{x}{x}{z}$
can be regarded as a pattern that a $t \in E$
may or may not conform to.
By filtering the multiplication relation's $\pair{\tau}{E}$
tuples through pattern $p$,
$\pair{\tau}{E}:p$ becomes, in its way, the squaring relation.

Let us now return to Example~\ref{ex:parentChild}
to see how filtering can be used here.
Suppose we have a set
$V = \{x,y,z \}$
and an index set
$I = \{\parent,\child \}$.
Let $p$ and $q$ both be in $I \rightarrow V$,
with $p$ such that
$p_\parent = x$, 
$p_\child =y$, 
and
with $q$ such that
$q_\parent = y$, 
$q_\child = z$. 

The relation
$$
\pi_{\{x,y\}}(\pc:p \;\Join\; \pc:q)
$$
is the equivalent of the relation resulting from the SQL query.

In this example, we have followed database usage in
making the index set
$I = \{\parent,\child \}$ non-numerical.
If we set $I = 2$,
then we have the convenient notation
$\pair{x}{y}$ for $p$
and
$\pair{y}{z}$ for $q$.
The query then becomes
\begin{equation} \label{eq:parentQuery}
\pi_{\{x,z\}}(\pc:\pair{x}{y} \;\Join\; \pc:\pair{y}{z}).
\end{equation}

With the relational operations defined so far (the set-like operations,
projection, cylindrification, join, and filtering)
we can define a wide variety of queries.
We do this by writing expressions
in the informal language of set theory.
This does not mean that a ``query language'' exists.
Of course, to make such expressions readable for a machine,
they have to be formalized.
Only then a ``language'' exists,
and then only in a technical sense.
Thus a ``query language'' should only arise
as part of the user manual of a software package.
It has no place in expositions of the relational data model.

Similarly, as soon as we have relations
and operations on them, an algebra exists.
But this is only so in a technical
and not in any substantial sense.
Here is an example of what I mean by the existence
of an algebra in a substantial sense.
For example, we could observe
that the set of integers is closed under addition,
that addition has an inverse,
and that zero is the neutral element under addition.
That we then have an algebra in a substantial sense
is borne out by the fact
that this algebra is a group
and that examples of groups exist
that don't look like the integers at all,
yet have certain interesting properties in common
that are expressed in the usual group axioms and theorems.

It happens to be the case that the set-like operations
together with cylindrification and certain special relations
like the diagonals constitute an algebra worth the name,
the cylindric set algebra.
The operations were identified by Tarski.
He only talked about an algebra when a significant
theorem  and interesting properties had been identified,
and then only in abstracts less than a half page long
\cite{trsk52,trskth52}.
Only much later, when significant algebraic results were obtained,
were cylindric algebras made the subject
of a longer publication \cite{hmt85}.

In the database world things work differently:
already in the first few years,
Codd proposed two query languages for the relational data model.
One, the relational calculus,
was to be of a nonprocedural nature,
so as to make it easy for users to relate the query
to their intuitive perception of the real-world situation
described by the database.
The other query language, the relational algebra,
is distinguished by operations on relations as algebraic objects.
This language was intended to facilitate query optimization.
However, for some decades the query language
used most widely in practice is SQL,
which has neither of these characteristics.

I have avoided introducing a relational algebra
or a query language.
Instead I have limited myself to introducing operations on relations:
the set-like operations, projection, cylindrification, join,
and filtering.
The closest I come to a query language are the expressions
in the informal language customary and universal
to all mathematical discussions.

In spite of these limitations,
it is striking how close a query such as the one
in Equation (\ref{eq:parentQuery}) comes to its equivalent
in Datalog \cite{mrwrrn88},
one of the query languages proposed in the literature.
In Datalog, this query would be
\begin{equation}
\answer(x,z) \leftarrow \pc(x,y), \pc(y,z).
\end{equation}
One of the advantages of Datalog over the relational calculus of Codd
is that Datalog is a subset of the first-order predicate calculus,
a relational calculus that antedates computers by half a century.
The predicate calculus fully embodies Codd's ideal of a
declarative relational language.
The reason Codd could not adopt it,
is that a relational algebra counterpart was not known
at the time.
Recent research \cite{hmt85,imlip84,vnmdn06} has remedied this deficiency.

\section{Conclusion}
When I advocate basic set theory for computer science,
I don't mean finding the right formula to copy.
Neither this paper, nor any of the books may have the right
formula.
For example, neither Bourbaki \cite{brbk39} nor Halmos \cite{hlm60}
give exactly the notion of relation required for databases.
The difference is mathematically trivial,
but crucial for the relational data model.
Neither \cite{brbk39}, nor \cite{hlm60}, nor this paper
can anticipate such trivial but crucial variations.

For the mathematicians consulted by Codd,
the distinction between $n = \{0,\ldots,n-1\}$ and more general index sets
in a relation was trivial, so they used the most familiar, which is $\iot{n}$.
But the failure to allow for general index sets
has continued to trouble the relational data model for years.
For Codd it was dangerous to know more than the general idea
that a relation is a set of tuples, that a tuple is a function,
and that arguments for a function don't have to be numbers.
And for us it is dangerous to copy the formulas from Codd.
We should remember his idea, and take it from there,
as best we can.

Hence the advice given by Halmos \cite{hlm60} ``Read this, and forget it.''
Or as Goethe said: ``Was du ererbt von deinen V\"{a}tern hast,
erwirb es um es zu besitzen.''\footnote{
What you have inherited from your fathers,
create it, so that it may be yours.
}

In other words, don't copy the formulas.

\section{Acknowledgements}
Many thanks to
Hajnal Andr\'eka,
Philip Kelly,
Michael Levy,
Belaid Moa,
Istv\'an Nemeti,
Alex Thomo,
and Bill Wadge
for helpful discussions and suggestions.


\end{document}